\documentclass[11pt,oneside,letterpaper]{article}
\usepackage{amssymb}
\usepackage{amsmath}
\usepackage[dvips]{graphicx}
\usepackage{setspace}
\usepackage{amsfonts}
\usepackage{fancyhdr}
\usepackage{xcolor}
\usepackage{graphicx}
\usepackage{rotating}
\usepackage{comment}
\usepackage{color}
\usepackage{cite}
\usepackage{braket}
\definecolor{darkgreen}{rgb}{0,0.5,0}
\definecolor{darkblue}{rgb}{0,0,0.6}
\definecolor{purple}{rgb}{0.4,.2,0.7}

\newcommand{\f}{\frac}

\newcommand{\be}{\begin{equation}}
\newcommand{\ee}{\end{equation}}

\usepackage[colorlinks=true,citecolor=darkgreen,linkcolor=black,urlcolor=purple]{hyperref}

\usepackage{pdfsync}
\makeatletter
\newcommand*{\defeq}{\mathrel{\rlap{%
                     \raisebox{0.3ex}{$\m@th\cdot$}}%
                     \raisebox{-0.3ex}{$\m@th\cdot$}}%
                     =} 
\makeatother

\def\be{\begin{eqnarray}}
\def\ee{\end{eqnarray}}

\newcommand{\bea}{\begin{eqnarray}}
\newcommand{\eea}{\end{eqnarray}}
\def\ben{\begin{equation}}
\def\een{\end{equation}}

\let\a=\alpha \let\b=\beta \let\g=\gamma  
   \let\k=\kappa
 \let\m=\mu    \let\r=\rho
\let\s=\sigma \let\t=\tau

\let\f=\frac

\def\be{\begin{equation}}
\def\ee{\end{equation}}
\def\ba{\begin{array}}
\def\ea{\end{array}}

\def\ba#1\ea{\begin{align}#1\end{align}}
\def\bs#1\es{\begin{split}#1\end{split}}

\allowdisplaybreaks  % allow page breaks in displayed eqs

\numberwithin{equation}{section}
\begin{document}
\onehalfspacing

\begin{center}

~
\vskip5mm

{\LARGE  {
The central dogma and cosmological horizons
\\
\ \\
}}

Edgar Shaghoulian

\vskip5mm
{\it David Rittenhouse Laboratory, University of Pennsylvania,\\
Philadelphia, PA 19104, USA
} 

\vskip5mm

{\tt eshag@sas.upenn.edu }

\end{center}

\vspace{4mm}

\begin{abstract}
\noindent
The central dogma of black hole physics -- which says that from the outside a black hole can be described in terms of a quantum system with exp$(\text{Area}/4G_N)$ states evolving unitarily -- has recently been supported by computations indicating that the interior of the black hole is encoded in the Hawking radiation of the exterior. In this paper, we probe whether such a dogma for cosmological horizons has any support from similar computations. The fact that the de Sitter bifurcation surface is a minimax surface (instead of a maximin surface) causes problems with this interpretation  when trying to import calculations analogous to the AdS case. This suggests anchoring extremal surfaces to the horizon itself, where we formulate a two-sided extremization prescription and find answers consistent with general expectations for a quantum theory of de Sitter space: vanishing total entropy, an entropy of $A/4G_N$ when restricting to a single static patch, an entropy of a subregion of the horizon which grows as the region size grows until an island-like transition at half the horizon size when the entanglement wedge becomes the entire static patch interior, and a de Sitter version of the Hartman-Maldacena transition.
 \end{abstract}
%\vspace{.2in}
%\vspace{.3in}

\pagebreak
\pagestyle{plain}

\setcounter{tocdepth}{2}
{}
\vfill

\ \vspace{-2cm}
\renewcommand{\baselinestretch}{1}\small
\tableofcontents
\renewcommand{\baselinestretch}{1.15}\normalsize

\section{Introduction}
The gravitational path integral has provided many powerful insights since its incarnation \cite{Gibbons:1976ue}. It seems both self-consistent and supported by various microscopic calculations, beginning with \cite{Strominger:1996sh}. It is not universally valid, motivating the conjecture that its application is correct if and only if the gravitational theory under consideration is holographic \cite{Harlow:2020bee}. If this is true, then one may hope that application of the gravitational path integral in the cosmological context  can provide clues about a holographic description. 

Recently, the central dogma of black hole physics -- the claim that a black hole can be described from the outside as a unitary quantum system with exp$(\text{Area}/4G)$ states -- has been supported by computations from the gravitational path integral \cite{Penington:2019kki, Almheiri:2019qdq}, building on \cite{Lewkowycz:2013nqa, Faulkner:2013ana, Dong:2016hjy, Dong:2017xht, Penington:2019npb, Almheiri:2019psf}. Extrapolating to the cosmological context, one might hope that the cosmological horizon can be described from the inside as a unitary quantum system with exp$(\text{Area}/4G)$ states \cite{Bousso:1999dw, Banks:2000fe, Fischler}; see \cite{Bousso:2000nf, Banks:2001yp, Banks:2002wr, Parikh:2002py, Dyson:2002nt, Banks:2003pt, Banks:2006rx, Banks:2015iya, Banks:2018ypk, Banks:2020zcr, Susskind:2021yvs} for a development of these ideas. 

This conjecture is on much shakier ground for multiple reasons. One reason is that there is little microscopic support for the Gibbons-Hawking entropy of cosmological horizons  \cite{Gibbons:1976ue} as coming from a Hilbert space of states (although see e.g. \cite{Bousso:2001mw, Alishahiha:2004md, Dong:2010pm, Anninos:2011af, Anninos:2017eib, Anninos:2017hhn, Anninos:2018svg, Dong:2018cuv, Gorbenko:2018oov, Geng:2019bnn, Anninos:2020geh, Muhlmann:2021clm} for holographic perspectives). Another reason is that the zero-point entropy, which is the piece that is interestingly fixed by the gravitational path integral, is more mysterious in the cosmological context. This is related to the fact that you can't create a cosmological horizon the way you can a black hole horizon. Finally, while it's natural that the cosmological horizon encodes the exterior (for example, when you throw something past the cosmological horizon, it grows) the way a black hole encodes the interior, that leaves open some puzzles. First, what encodes the interior of the cosmological horizon?\footnote{One way to try to address this is to embed a de Sitter horizon in a spacetime with a timelike boundary \cite{Anninos:2018svg, Anninos:2017hhn, Mirbabayi:2020grb, Fischetti:2014uxa, Freivogel:2005qh}.} Second, different observers' exteriors can overlap, leading to redundant encoding of a nature not encountered in AdS/CFT; it would be akin to an observer in the right exterior of the eternal black hole being able to decode the interior without access to any of the left exterior. How is this supposed to be reconciled?

In this paper we will focus on de Sitter spacetime and do various computations of entanglement entropy to see if the exterior of the horizon is encoded in the interior. Doing something precisely analogous to the black hole context is difficult because we don't naturally have an asymptotic region where gravity is weak and we have infinite volume.\footnote{In this sense,  computations of entropy at $\mathcal{I}^+$ \cite{Chen:2020tes, Hartman:2020khs}, or after exiting from inflation, are more analogous to the black hole computations. Such reasoning also motivates the introduction of a ``census taker" \cite{Freivogel:2006xu, Susskind:2007pv}.} Physically, the de Sitter horizon reabsorbs its own Hawking radiation, and we can only collect so much of it without backreacting on the spacetime. And the typical wavelength is of the size of the observable universe. Practically, we will nevertheless assume we can ``freeze" gravity far from the de Sitter horizon, allowing us to define regions which do not fluctuate and whose entropy we will compute. The results do not support a picture of the interior encoding the exterior and instead suggest anchoring extremal surfaces to the horizon itself, which we will also explore.

\subsection*{{\it Summary}}
We use a conformal map between de Sitter and flat spacetime to constrain computations of entanglement entropy in de Sitter. When we freeze gravity in the interior of a piece of the static patch (and a similar antipodal region), we will show that generalized entropy computations in de Sitter space have  a natural quantum extremal surface \cite{Faulkner:2013ana, Engelhardt:2014gca}, stemming from the classical extremal surface (the bifurcation surface) \cite{Ryu:2006bv, Hubeny:2007xt}, which is a minimum in time and a maximum in space, i.e. a minimax surface instead of a maximin surface. This leads to violations of entanglement wedge nesting, a basic consistency principle of any such picture. The only other extremal surface is the degenerate one which sits right on the boundary between the frozen and unfrozen regions. This leads us to considering anchoring extremal surfaces on the horizon itself, for which we formulate a two-sided extremization procedure inspired by a similar situation in AdS/CFT. This gives sensible results:

\begin{itemize}
\item The entropy of a subregion of one horizon theory grows with twice the size of the subregion. This threatens the bound set by a quantum system with exp$(A/4G)$ states, but a transition occurs at half the horizon size and we remain pegged at $S = A/4G$ as the region size increases. The entanglement wedge transitions from none of the interior to the entire interior at this point. These features mimic the island transition in an eternal black hole \cite{Penington:2019npb, Almheiri:2019psf, Almheiri:2019yqk}.
\item The entropy of the union of the two horizons vanishes.
\item The entropy of similar cuts on each horizon has a transition like that of Hartman-Maldacena \cite{Hartman:2013qma}, switching between a connected surface which goes through the inflating region and a disconnected one which sits on the two horizons. This transition stops the entanglement from exceeding the amount allowed by a horizon theory with exp$(A/4G)$ states.
\item Entanglement wedge nesting is satisfied. 
\end{itemize}
The nature of the horizon theory -- in particular, its locality properties and coupling to gravity, how it depends on the choice of horizons, and how it responds to the dynamical nature of the horizon -- will not be addressed in this work. 

Entanglement entropy in de Sitter space based on horizon-anchoring was recently discussed in \cite{Susskind:2021dfc, Susskind:2021esx} (see also \cite{Sanches:2016sxy, Nomura:2017fyh} for previous work on anchoring to the horizon).\footnote{It was also discussed by the present author at a seminar at MIT in March.} Although the framework in \cite{Susskind:2021dfc, Susskind:2021esx} was not explicitly two-sided extremization the results appear similar.

\vspace{4mm}
\section{CFT entropy in de Sitter spacetime}\label{monotonic}
In this section we will use the monotonicity of matter entropy to derive some constraints on the CFT entropy in de Sitter spacetime by conformally mapping to the plane. An annulus in the plane will map to an angular section $\theta \in (\theta_i, \theta_f)$ on the spatial sphere of the $t=0$ time slice. This is the entropy we want to compute. The constraints on the annulus entropy have been worked out in e.g. \cite{Hirata:2006jx}. 

\subsection{Monotonicity of matter entropy of annulus in plane}\label{secineq}
Let's begin by considering the plane, $ds^2 = d\r^2 + \r^2 d\Omega_{d-2}^2$, and radii $\r_0 < \r_1 < \r_2<\r_3$. We will consider the quantum state on the entire plane to be pure. The entanglement entropy for an annulus between $\r \in (\r_i,\r_j)$ will be denoted $S(\r_i, \r_j)$. Strong subadditivity $S(A) + S(B) \geq S(A \cup B) + S(A\cap B)$ says
\be\label{SSA}
S(\r_0,\r_2)+S(\r_1,\r_3)\geq S(\r_1,\r_2)+S(\r_0,\r_3)
\ee
If we take the limit $\r_3 \rightarrow \infty$ we get
\be
S(\r_0, \r_2) \geq S(\r_1, \r_2)+S(0,\r_0)-S(0,\r_1)\,,
\ee
while if we take $\r_0 \rightarrow 0$ we get
\be
S(\r_1, \r_2)\leq S(\r_1,\r_3)+S(0, \r_2)-S(0, \r_3)\,.
\ee
So far this is true for an arbitrary QFT$_d$, but for the rest of this section we consider a CFT$_{d>2}$. Defining the finite part of the entropy as $S_0$, we have $S_0(0, \r_i) = S_0(0, \r_j)$ for the sphere entanglement entropy and  $S_0(\r_i , \r_j) = S_0(\r_j/\r_i)$.\footnote{Recall that the finite part of the entropy in even dimensions is scheme-dependent due to pollution from the logarithmic divergence; we assume a position-independent scheme so that this constant ambiguity falls out of the derivative relations below. In particular, the finite piece is defined so that it does not include any terms like $\log \rho_i$, which will instead make an appearance in the full entropy in the next subsection.}
Since all divergences cancel in these expressions, we get
\be\label{genineq}
  S_0( \r_2/\r_1) \leq S_0(\r_2/\rho_0)\,,\qquad S_0(\r_2/\r_1)\leq S_0(\r_3/\r_1)
\ee
 Defining $\r_j / \r_i = R \geq 1$ gives 
\begin{align}
S_0'(R) \geq 0\,
\end{align}
for the annulus entanglement entropy. In other words, the entanglement entropy of the annular region increases as you thicken the annulus.

We can also get a constraint on the second derivative of the entropy of the annulus. Take $\r_3 =\r_2(1+\epsilon/R)$, $\r_0 = \r_1(1-\epsilon/R)$ with $R = \r_2/\r_1$ to rewrite \eqref{SSA} as
\begin{align}
S_0(R+2\epsilon)-2S_0(R+\epsilon)+S_0(R) \propto S_0''(R) \leq 0\,.
\end{align}
We also have that 
\be\label{ffactor}
S_0(R\rightarrow 1) \approx -\k \f{\text{Area}(S^{d-2})}{(R-1)^{d-2}}\,,\qquad S_0(R\rightarrow \infty) \approx -2F\,,\qquad \k > 0.
\ee
The former exression comes from approximating the thin annulus as a parallel plate geometry with plates of area Area$(S^{d-2})\rho_2^{d-2}$ and separation $\r_2-\r_1$. The sign of $\k$ is fixed by the constraint $S_0'(R) \geq 0$. The latter expression comes from computing the entropy in the complement region, which should factorize into the sum of the entropies of two spheres of radii $\r_1$ and $\r_2$, each of which contributes $-F$. In three dimensions $F$ is related to the $F$-theorem and we have $F > 0$. In even dimensions the quantity (including its sign) is scheme-dependent. 

Altogether, we see that the function is monotonically non-decreasing from $-\infty$ to $-2F$ and remains concave throughout. In particular, the first derivative decreases from $+\infty$ all the way to zero, by continuity taking all values inbetween. 

\subsection{Generalized entropy}
In our applications we will want to consider the generalized entropy:
\be
S_{\rm gen} = \f{{\rm Area}}{4G} + S_{\rm CFT}\,.
\ee
In three and four dimensions the full entropy for an annular region is given by
\be
S_{\rm CFT}^{(3d)} = \g \f{2\pi (\r_1+\r_2)}{\epsilon} + S_0^{(3d)}\,,\qquad S_{\rm CFT}^{(4d)} = \g \f{4\pi (\r_1^2+\r_2^2)}{\epsilon^2} -4a \log \f{\r_1\r_2}{\epsilon^2} + S_0^{(4d)}\,.
\ee
The renormalization of Newton's constant leads to divergences that cancel against the ones in the matter entropy, and the generalized entropy is expected to be a finite quantity:
\be
S_{\rm gen}^{(3d)} = \f{{\rm Area}}{4G} + S_0^{(3d)}\,,\qquad S_{\rm gen}^{(4d)} = \f{{\rm Area}}{4G} -4a \log (\r_1\r_2)+S_0^{(4d)}\,.
\ee
We define $\widehat{S}$ as
\be
\widehat{S}^{(3d)}=S_0^{(3d)}\,,\qquad \widehat{S}^{(4d)}=-4a\log(\r_1\r_2) + S_0^{(4d)}\,.
\ee
It can heuristically be thought of as the ``matter" part of the generalized entropy, although the split into the matter piece and the gravitational piece is not meaningful, since only the combination is finite. $\widehat{S}^{(3d)}$ is monotonic and concave from the relations in section \ref{secineq}, while $\widehat{S}^{(4d)}$ has the anomaly  term to deal with. As $\r_1 \rightarrow \r_2$ the anomaly piece contributes a finite amount, so we still have $\widehat{S}^{(4d)}(R\rightarrow 1) \approx S_0^{(4d)}(R\rightarrow 1)$. On the other hand, due to the anomaly, the behavior under thickening the annulus depends on whether we make $\r_1$ smaller or $\r_2$ bigger. In our applications we will be concerned with making $\r_1$ smaller, in which case we have 
\be
-\partial_{\r_1}\widehat{S}^{(4d)}(\r_1,\r_2) \geq 0\,,\qquad \partial_{\r_1}^2\widehat{S}^{(4d)}(\r_1,\r_2) = \f{4a}{\r_1^2} + S_0''(R)\,.
\ee
So the sign of the second derivative is not fixed. 

\subsection{Mapping to de Sitter}\label{dsmonotonic}
\begin{figure}
  \centering
  \includegraphics[scale = .08]{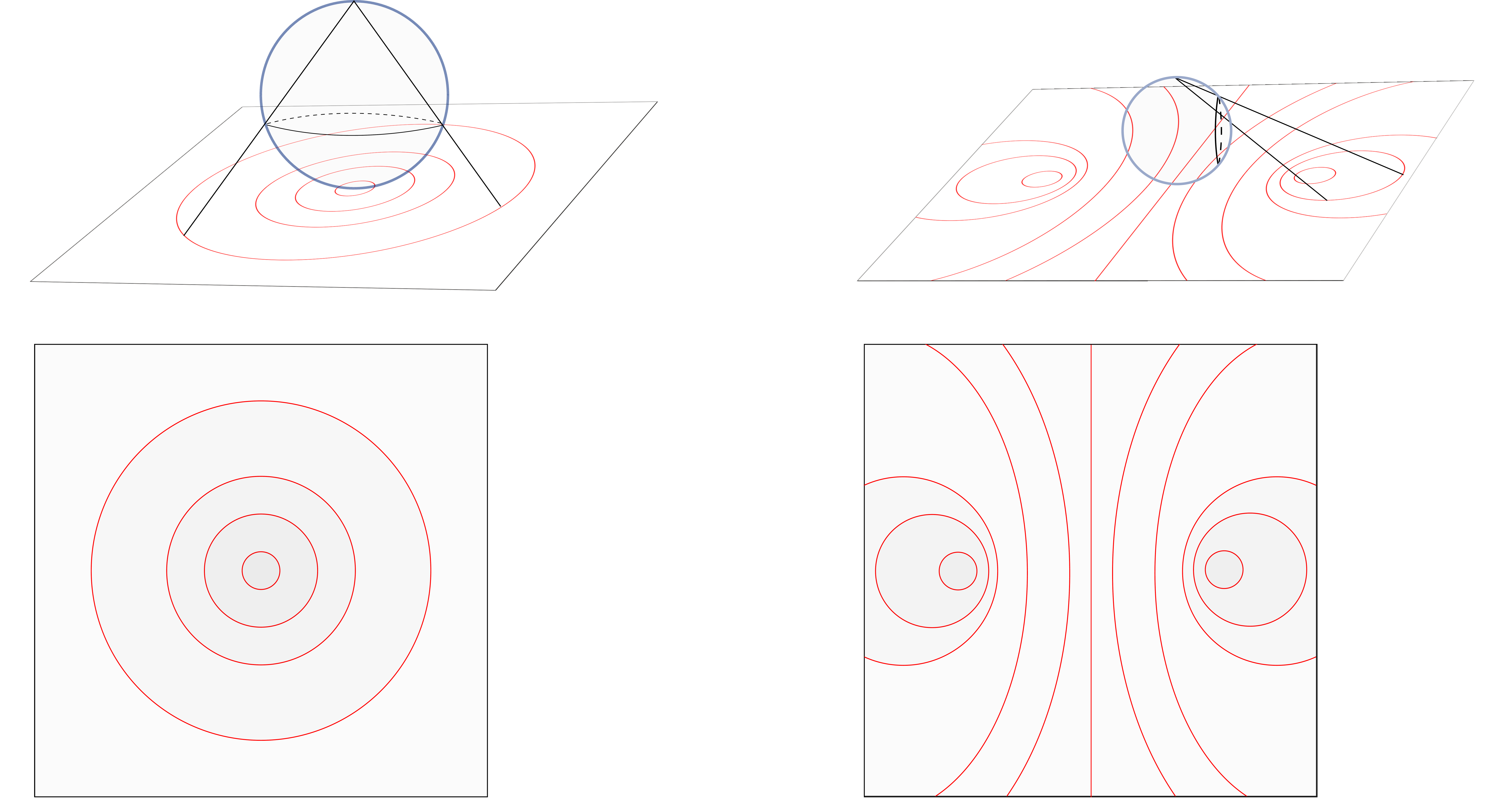}
\caption{Two different stereographic projections. The one on the left foliates the sphere by lower-dimensional spheres from north to south, which get projected onto the plane as spheres of increasing radius. The one on the right foliates the sphere by lower-dimensional spheres from west to east, which get projected onto the plane as two sets of spheres surrounding two points. The sphere passing through the north and south poles gets mapped to a lower-dimensional plane. The bottom line shows a direct view of the parameterization of the plane, radial coordinates on the left and owl coordinates on the right.}\label{stereographics}
\end{figure}

Let's say we want to compute the entropy on de Sitter spacetime, 
\be
ds^2 = -dt^2 + \cosh^2 t(d\theta^2 + \sin^2 \theta d\Omega_{d-2}^2)\,,\qquad t\in (-\infty, \infty), \quad \theta \in (0,\pi)
\ee 
for a belt-like region $\theta\in (\theta_1, \theta_2)$ at $t=0$. Since we are at $t=0$ we can do a Euclidean computation on the sphere $ds^2 = d\t^2 + \cos^2 \t(d\theta^2 + \sin^2 \theta d\Omega_{d-2}^2)$ at $\t = 0$. In Euclidean signature the full sphere is conformally related to the plane, as can be seen by the stereographic projection:
\be\label{weyl}
ds^2 = \f{4 dx_i^2}{(1+\sum_{i=1}^d x_i^2)^2} = \Omega^{-2} dx_i^2\,,\qquad \Omega = \f{1+\sum_{i=1}^d x_i^2}{2}\,.
\ee
A natural way to interpret this stereographic projection is to have Euclidean time $\t$ run from the south pole to the north pole. Then the $(d-1)$-spheres foliated by $\t$ map to $(d-1)$-spheres foliated in the plane by radius. This projects the south pole to the origin and the north pole to infinity, see the left-hand-side of figure \ref{stereographics}. 

Another interpretation is to have Euclidean time $\t$ run from the ``west" pole to the ``east" pole, while keeping the stereographic projection fixed, see the right-hand-side of Figure \ref{stereographics}. The rest of the stereographic projection we interpret in the earlier fashion: the $(d-2)$-dimensional spheres on our $\t = 0$ $(d-1)$-dimensional sphere map to $(d-2)$-dimensional spheres in the Euclidean plane foliated by radius. Thus, a belt like region $\theta \in (\theta_1, \theta_2)$ maps to an annular region $\r \in (\r_1, \r_2)$ in the plane. To relate these quantities, consider de Sitter spacetime in static patch coordinates:
\be
ds^2 = -(1-r^2)dt^2 + \f{dr^2}{1-r^2} + r^2 d\Omega_{d-2}^2 = -\cos^2 \theta dt^2 + d\theta^2 + \sin^2 \theta d\Omega_{d-2}^2
\ee
In the latter coordinates, where $r = \sin \theta$, we see that $\theta \in [0,\pi/2]$ covers one static patch, although we can continue to $\theta \in (\pi/2, \pi]$ to enter the neighboring static patch.

In these coordinates, a belt-like region $r \in (r_1, r_2)$ maps into an annular region $\r \in (\r_1, \r_2)$, with the relation between the two being given by the Weyl factor (or some trigonometry)
\be\label{flatds}
r = \f{2\r}{1+\r^2}\,.
\ee

\section{Central dogma for cosmological horizons}
Consider de Sitter spacetime again in static patch coordinates,
\be
ds^2 = -(1-r^2)dt^2 + \f{dr^2}{1-r^2} + r^2 d\Omega_{d-2}^2 \,.
\ee
This spactime has a horizon at $r=1$, which has a thermodynamic entropy given by 
\be
S = \f{{\rm Area}(S^{d-2})}{4G}\,,
\ee
similar to the thermodynamic entropy of black holes. 

\begin{figure}
  \centering
  \includegraphics[scale = .2]{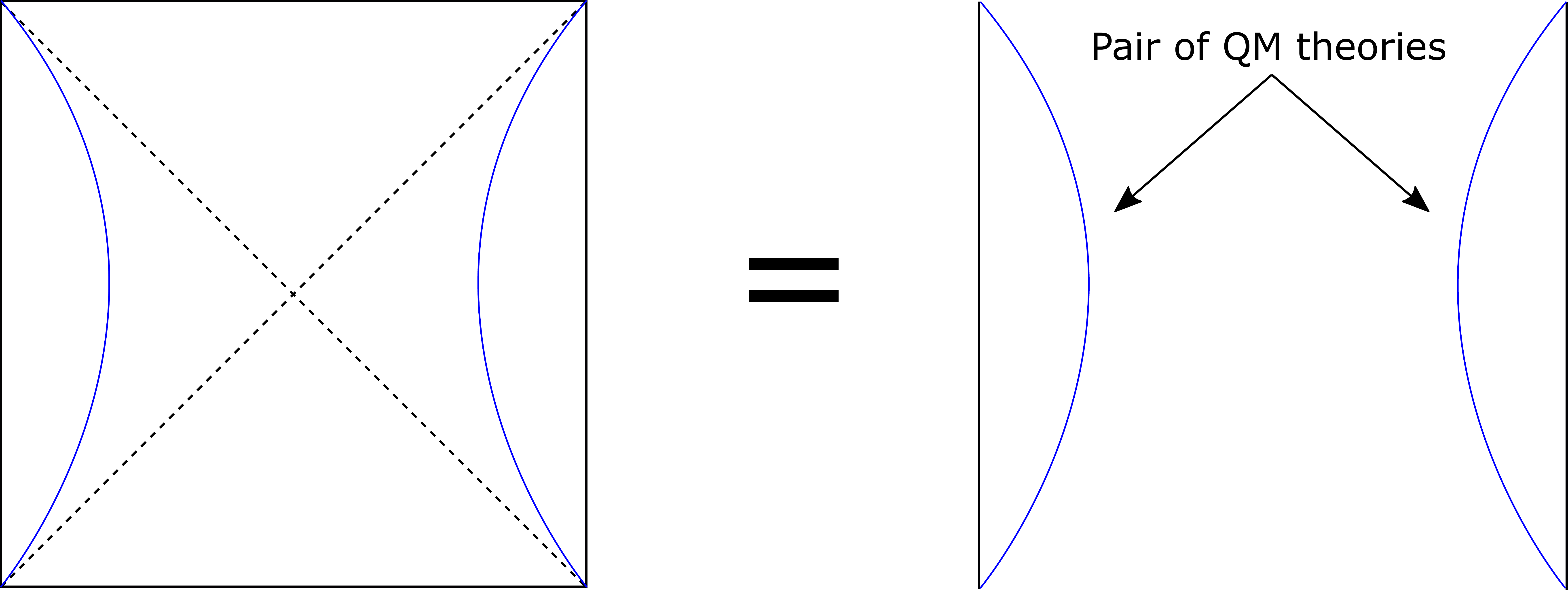}
\caption{The de Sitter region is replaced by a pair of quantum-mechanical theories, each described by a Hilbert space of $e^{A/4G}$ degrees of freedom.}\label{dSquantum}
\end{figure}

For the ensuing analysis we will consider an observer limited to some region $r < r_c$, with $r_c < 1$, where we freeze gravity.\footnote{As discussed in the introduction it is not clear that this is a sensible approximation, although the general lesson regarding the minimax surface will apply in more reasonable spacetimes where we can freeze gravity, e.g. embedding de Sitter in asymptotically AdS spacetime or after exiting inflation.} In analogy to the black hole case we will refer to this region as the ``bath" \cite{Rocha:2008fe, Anninos:2019oka, Almheiri:2019psf}. We will assume that the de Sitter horizon, as viewed from the observer's vantage point, can be thought of as a quantum system with $e^{A/(4G)}$ degrees of freedom. This is represented in figure \ref{dSquantum}, where the blue curves are at $r= r_c$. In this exact quantum-mechanical description, we would like to compute the entropy of one of the quantum-mechanical theories plus a small part of the bath region at time $t=0$. We will be using the QES prescription \cite{Ryu:2006bv, Hubeny:2007xt, Faulkner:2013ana, Engelhardt:2014gca}, which requires extending the region into the bulk and extremizing the generalized entropy with respect to the endpoint of the region, see the right-hand-side of Figure \ref{ds3}. Interestingly, we can show that there is a nontrivial quantum extremal surface for this computation.

Before exhibiting the quantum extremal surface, let's consider some aspects of the classical problem. There is an obvious classical extremal surface, which is simply the bifurcation point in the Penrose diagram. This point lies at $t=0$ and therefore does not spontaneously break the $\mathbb{Z}_2$ time-reflection symmetry of the problem. Notice that this surface is quite peculiar as compared to surfaces usually considered in these problems: it is maximal in space and minimal in time. These means that in the Euclidean computation it will be a maximum of the action instead of a minimum. Assuming that it contributes to the path integral, it implies that the entropy of our quantum-mechanical theory is given precisely by $A/(4G)$. This is simply the Gibbons-Hawking answer.

Now we consider the quantum version of the problem. First let's remember what happens in the case of an eternal black hole. The quantum extremal surface lives slightly outside the horizon. This is because the area is decreasing as the horizon is approached, while the matter entropy is increasing. At the bifurcation point the first derivative of the area vanishes, so the two terms must balance somewhere before this point is reached. Thus, the quantum extremal surface is in the right asymptotic region. For de Sitter spacetime, as mentioned above, the area \emph{increases} as the horizon is approached, which will make a profound difference below.

We consider Einstein gravity with a positive cosmological constant coupled to CFT$_d$. We consider the Hartle-Hawking state for the matter on dS$_d$. The region $\tilde{I}$ is a belt-like region on the sphere $S^{d-1}$ at time $t=0$, with left and right endpoints given by $r_1$ and $r_2$, respectively. By restricting to $t=0$ we are assuming the $\mathbb{Z}_2$ time-reflection symmetry is not spontaneously broken. The generalized entropy for this region is
\be
S_{\rm gen} = \f{{\rm Area}(S^{d-2})r_1^{d-2}}{4G} + \widehat{S}_{\rm mat}(\tilde{I})\,.
\ee
We now restrict to $d=3$ and $d=4$.
\subsection{dS$_3$}
\begin{figure}
  \centering
  \includegraphics[scale = .2]{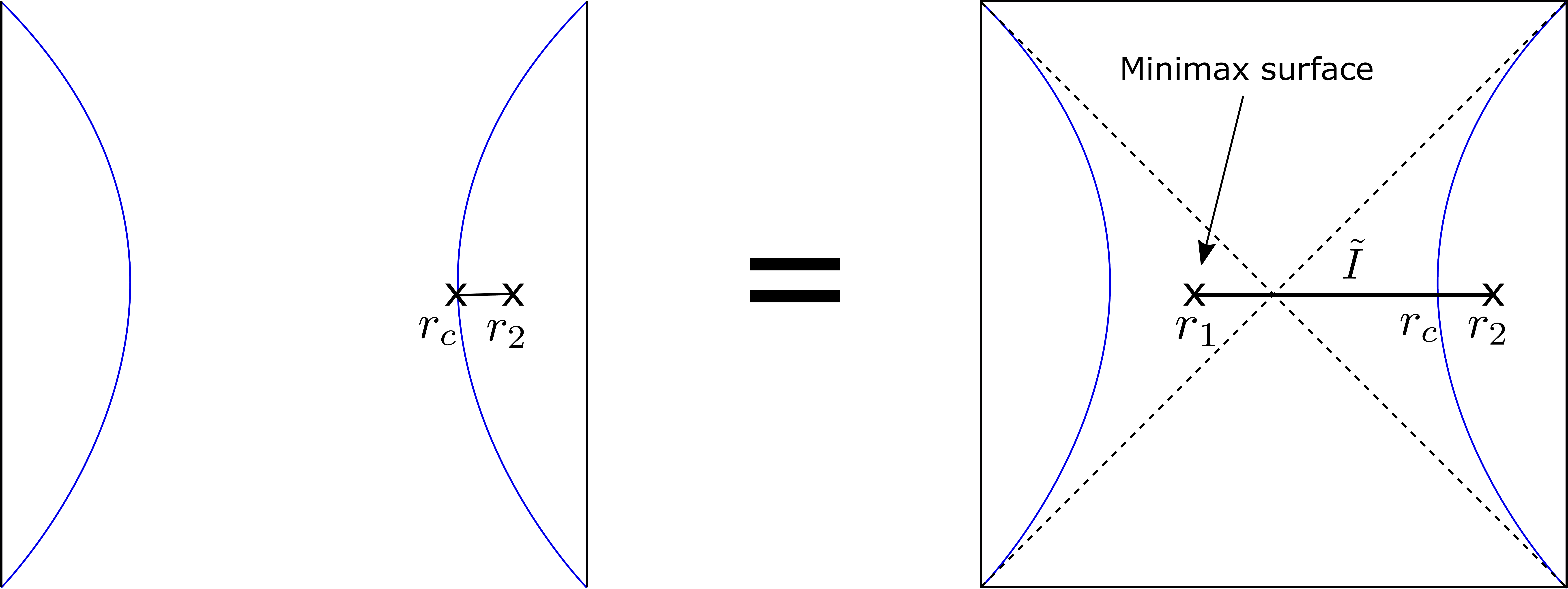}
\caption{The fine-grained entropy for regions specified in the exact quantum-mechanical description (left side of equality) would equal the generalized entropy for the regions specified in the semiclassical spacetime (right side of equality) if the minimax surface shown was an acceptable quantum extremal surface.}\label{ds3}
\end{figure}

In three dimensions there is no conformal anomaly, so we have 
\be
\widehat{S}_{\rm mat}(\tilde{I}) = S_0(\r_2/\r_1)\,,
\ee
where $S_0(\r_2/\r_1)$ is the finite piece of the entanglement entropy in the plane between circles of radii $\r_1$ and $\r_2$, as discussed in section \ref{monotonic}. Thus the generalized entropy is
\be
S_{\rm gen} = \f{\pi r_1}{2G} +S_0\left(\r_2/\r_1\right),
\ee
where we can use \eqref{flatds} to go between the coordinates on the sphere and the plane. 
Let's say $r_2$ is chosen such that there exists some $r_f > r_c$ in the left static patch for which the annulus entanglement entropy $S_0(\r_2/\r_f)$ factorizes into approximately $-2F$ (see equation \eqref{ffactor}). This means there exists some $r_1$ in the left static patch for which we have an extremum with respect to an $r_1$ variation, see Figure \ref{ds3}. This is because the $r_1$ derivative of the gravitational entropy is positive at $r_f$ while the derivative of the matter entropy vanishes. The derivative of the gravitational entropy approaches zero as $r_1 \rightarrow 1$, i.e. as the horizon is approached, while the $r_1$ derivative of the matter entropy is non-positive as it makes the annulus thinner. This means at some point as the horizon is approached the two terms will have equal magnitude and opposite signs. The second derivative of the gravitational and matter entropies are both non-positive, indicating that the extremum is a maximum in space.

By the monotonicity of the matter entropy we also know there cannot be an extremum in the right static patch, since the sign of the derivative of the gravitational entropy flips and matches the sign of the derivative of the matter entropy. 

\subsection{dS$_4$}
In four dimensions we have 
\be
\widehat{S}_{\rm mat}(\tilde{I}) = S_0(\r_2/\r_1)-4a \log \f{2 \r_1}{1+\r_1^2}-4a\log \f{2\r_2}{1+\r_2^2}\,,
\ee
where the logarithmic terms have picked up a contribution from the trace anomaly, $\epsilon \rightarrow \Omega \epsilon$ with $\Omega$ given in \eqref{weyl}. So we have
\be
\f{\partial}{\partial \r_1} \widehat{S}_{\rm mat}(\tilde{I}) = \f{\partial}{\partial \r_1} S_0(\r_2/\r_1)-4a\f{1-\r_1^2}{\r_1+\r_1^3}\,.
\ee
The sign of the first term is negative by monotonicity and matches the sign of the second term for $\r_1 <1$, i.e. if we are in the left static patch. But we cannot use the same trick as above, since starting off in the ``factorized" channel where $\partial_{\rho_1} S_0(\rho_2/\rho_1) = 0$, the derivative of the matter entropy can still have a large negative contribution from the anomaly term. Thus it is possible that the derivative of the matter entropy is bounded above by some negative constant $-k$, while the derivative of the gravitational entropy is bounded above by $O(1)/G_N$. If the parameters are such that $k > O(1)/G$ then there will not be an extremum. 

We can avoid these issues by making the matter entropy a small perturbation on top of the gravitational entropy by taking $ a \ll 1/G$, so that a quantum extremal surface continues to exist (its precise location will not be important to us in the following).

\subsection{Inconsistency}
A basic consistency check on the region defined by the quantum extremal surface is entanglement wedge nesting. In other words, as we move $r_2$ to the right, we should find that $r_1$ moves to the left. This is actually false for our putative QES above, which is related to the fact that it is a minimax (minimum in time, maximum in space) surface instead of a maximin surface.

 Let's consider this first in 3d. We begin with the QES at some point in the left static patch. Moving $r_2$ to the right decreases the first derivative of the matter entropy; it cannot result in moving $r_1$ to the left because this increases both the first derivative of the matter entropy (decreasing its magnitude, since it is negative) and of the gravitational entropy, so they can no longer cancel.
Another way to think about it is to move $r_2$ so far to the right that the EE factorizes for $r_1$ in the right static patch. This means the $r_1$ derivative of the matter entropy vanishes, and will balance against the $r_1$ derivative of the gravitational entropy right at the bifurcation surface. In general dimension dS$_{d>3}$ we can run the same argument; in even dimensions the part of the anomaly term from $\rho_1$ satisfies $\partial_{\rho_1}S_{anom} = 0$ at the bifurcation surface, so we reach the same conclusion. 

We can check these expectations explicitly in JT gravity in dS$_2$ coupled to a CFT$_2$ where we can compute the entanglement entropy analytically. The solution we will consider is global dS obtained by an s-wave reduction of pure gravity in dS$_3$. We find
\be
x^{\pm}_1 \sim \f{k}{x^{\mp}_2}\,
\ee
for the extremal surface, with $k>0$ a semiclassical expansion parameter and $x^\pm = t \pm x$. The bifurcation point is $x^{\pm} = 0$, so $x_1$ lives in the left static patch. We increase the size of the interval by increasing $x^+_2$ and decreasing $x^-_2$. This leads to $x^+_1$ increasing and $x^-_1$ decreasing, i.e. the QES also moves to the right.

A related problem with the QES reaching into the neighboring static patch is that the purported island region of the right system will overlap with the island region of the left system. These issues are also related to the fact that perturbations make the dS Penrose diagram grow taller \cite{Gao:2001ut, Leblond:2002ns}. At a minimum this means these two systems should be thought of as interacting. Even interacting holographic systems, however, should obey entanglement wedge nesting, which is violated in this setup.

The root of these various problems seems to be the inclusion of \emph{minimax} surfaces in our entanglement entropy prescription. In the context of AdS/CFT, the surfaces which compute entanglement entropy and give reasonable behavior are always maximin surfaces. In particular, assuming the quantum focusing conjecture, entanglement wedge nesting, and strong subadditivity imply that the extremal surface is maximal in time \cite{Hartman:2020khs}, so it is not a surprise that entanglement wedge nesting is violated for our minimax surface.\footnote{Other arguments showing that the quantum focusing conjecture implies entanglement wedge nesting assume asymptotically AdS boundary conditions or a maximin prescription as the definition of the extremal surface (or both) \cite{Wall:2012uf, Akers:2016ugt}, so our result does not imply any problem for quantum focusing.} Surfaces which are maximal in space have instead recently been interpreted as being related to restricted complexity \cite{Brown:2019rox}. That is one possible role of the de Sitter bifurcation surface, although the covariantization of maximal surfaces would have to be generalized to include miniminimax surfaces analogous to the maximinimax surfaces of \cite{Brown:2019rox}.\footnote{The maximinimax prescription of \cite{Brown:2019rox} is that on every Cauchy slice you perform a minimax procedure, which is to sweep through all spatial slicings of the Cauchy slice, for each spatial slicing find the maximal area surface, and then minimize this over all spatial slicings. Finally, you maximize this over all Cauchy slices. The minimax part of the procedure is inspired by complexity in tensor networks. The maximization in time is not something coming from tensor networks but instead provides a covariantization which gives a nontrivial surface in e.g. the case of evaporating black holes, where a minimization would run into the singularity. Depending on how one sets up the problem in de Sitter spacetime, both miniminimax surfaces and maximinimax surfaces can give nontrivial answers.}

\begin{figure}
  \centering
  \includegraphics[scale = .2]{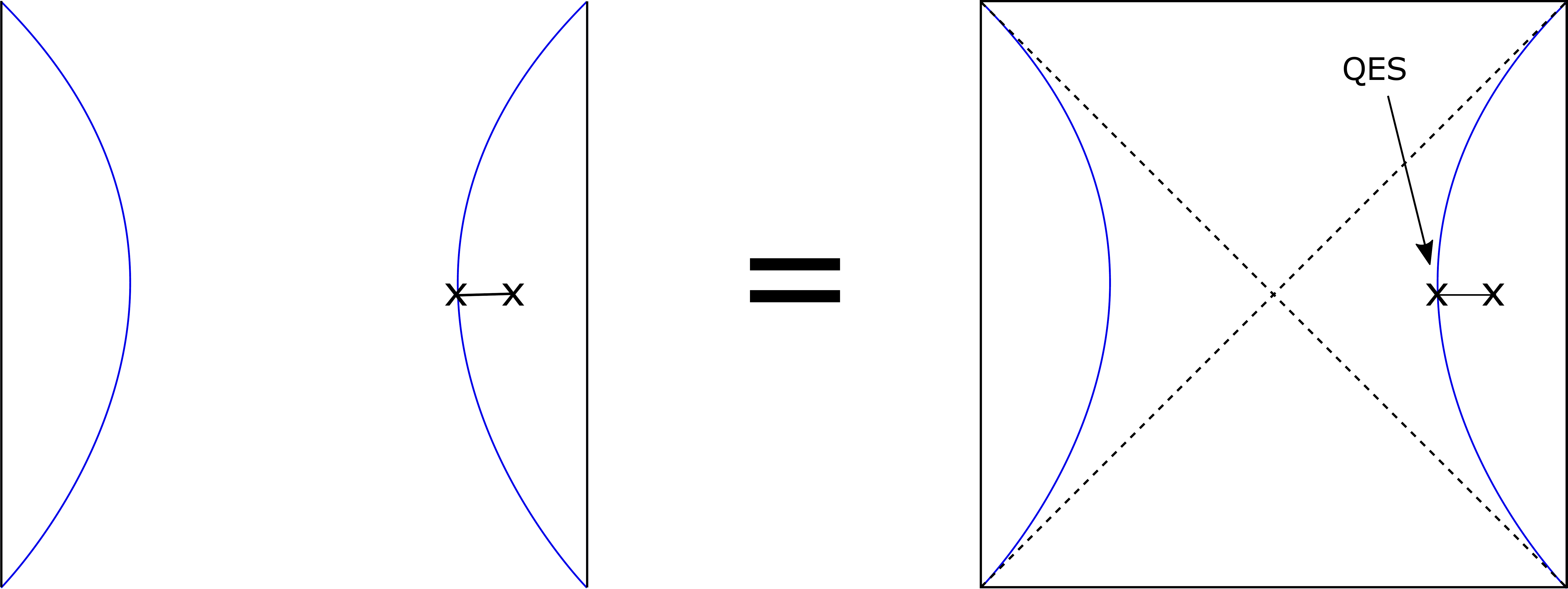}
\caption{The QES is given as the degenerate surface which sits right on the cutoff surface.}\label{dSdegen}
\end{figure}

If we exclude this surface, then the only remaining possible surface is the degenerate one which sits right at $r_c$ as in Figure \ref{dSdegen}. It is degenerate in the sense that if we minimize then the surface runs to the boundary and can't go any further due to the boundary conditions on the problem. In our problem $r_c$ was somewhat arbitrary, so the value implied for the entropy inherits this ambiguity. In some sense, this is the right answer, but it is not very useful. A natural $r_c$ to pick is the horizon, in which case the degenerate minimal surface ends up landing on the horizon itself, giving $S = A/4G$, a satisfying answer. Motivated by this, we will consider a slightly different prescription in the next section, where we anchor extremal surfaces to the horizon itself and consider a two-sided extremization.

\section{Anchoring on the horizon}
\begin{figure}
  \centering
 \hspace{-5mm} \includegraphics[scale = .13]{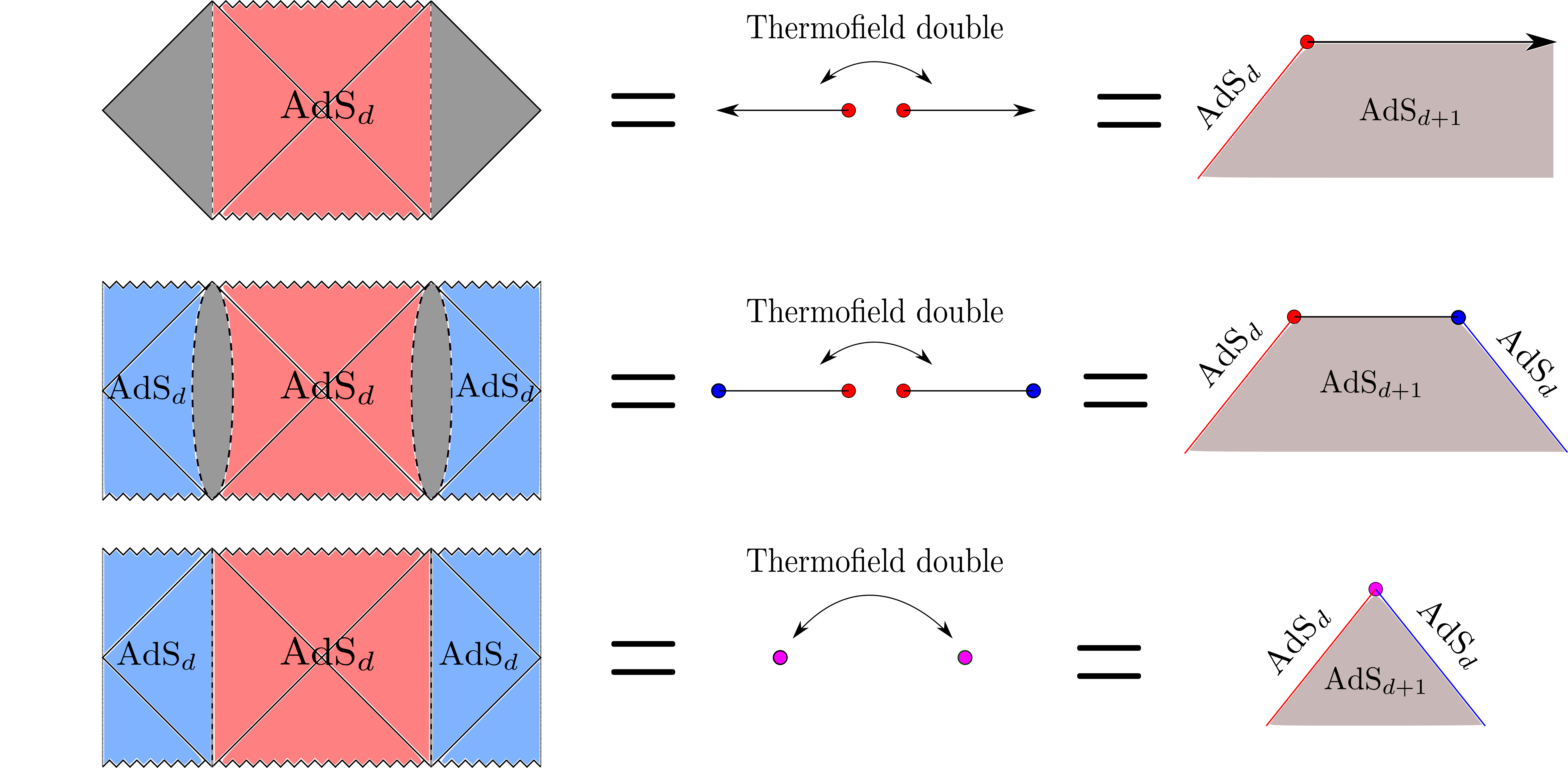}
\caption{We represent wedge holography as a series of modifications to the familiar example of the top line \cite{Almheiri:2019yqk,
 Almheiri:2019psy, Almheiri:2019qdq, Chen:2020uac, Chen:2020hmv}. On the top left we have the spacetime for the eternal black hole in AdS$_d$ coupled to two flat space baths. There is a holographic CFT$_d$ everywhere in the spacetime. The top middle represents the microscopic description, where we have a holographic CFT$_{d-1}$ (represented by a red dot) coupled to a $d$-dimensional bath hosting the holographic CFT$_d$, and this joint system's thermofield double partner. The top right illustrates the emergent $(d+1)$-dimensional description (the thermofield double part is not represented for simplicity). In the middle line we have the same representations, except we now have a finite bath and an additional holographic CFT$_{d-1}$ (represented by a blue dot) which is also coupled to the bath CFT$_d$ (and their thermofield double partners) \cite{Geng:2021iyq, Balasubramanian:2021xcm}. The Penrose diagram is identified at the left and right. The final line shrinks the bath size to zero, so the interacting CFT's sit right on top of one another \cite{Penington:2019kki, Geng:2020fxl}.  Notice in this case the boundary dual has spacetime to both sides. In our de Sitter analogy this will be the primary feature we are interested in, as our ``boundary dual" is on the cosmic horizon and has spacetime to both sides. The analog of the doubly emergent AdS$_{d+1}$ will not play a role. }\label{wedge}
\end{figure}

Placing a holographic dual to de Sitter space on the horizon itself has a long precedent \cite{Bousso:1999cb, Banks:2000fe, Alishahiha:2004md, Sanches:2016sxy, Banks:2020zcr, Anninos:2021ihe, Susskind:2021omt}.\footnote{The other natural place is $\mathcal{I}^+/\mathcal{I}^-$ \cite{Witten:2001kn, Strominger:2001pn, Maldacena:2002vr, Anninos:2011ui}; see \cite{Anninos:2012qw} for a review of various approaches and e.g. \cite{Narayan:2015vda, Narayan:2017xca, Narayan:2020nsc} for anchoring extremal surfaces to these boundaries.} There are serious questions about such a construction which we will not address, foremost being the dynamical nature of the horizon and the properties of the theory on the horizon: gravity does not obviously decouple there, so it seems the holographic dual would itself be gravitational, as in \cite{Alishahiha:2004md}. In the rest of this work we will compute entanglement entropy for cuts of the horizon as if it were a (possibly nonlocal) quantum system. In computing entanglement entropy, however, we are faced with another issue, which is that there is spacetime to both sides of the horizon. A similar issue has been explored in AdS/CFT, using Karch-Randall branes \cite{Randall:1999ee, Karch:2000ct} in what has been coined ``wedge holography" \cite{Mollabashi:2014qfa, Miao:2020oey, Akal:2020wfl, Bousso:2020kmy, Geng:2020fxl}. In the context of wedge holography, one has a $(d-1)$-dimensional CFT which is dual to a ``wedge" in AdS$_{d+1}$ bounded by two gravitating branes. However, another description is that the $(d-1)$-dimensional CFT is dual to two $d-$dimensional CFTs coupled to gravity. See Figure \ref{wedge}. These two systems are on either side of the $(d-1)$-dimensional CFT. The prescription for computing entanglement entropy for a subregion $A$ of the $(d-1)$-dimensional CFT is to find all $(d-1)$-dimensional extremal surfaces in the $(d+1)$-dimensional wedge whose boundary contains $\partial A$, which is $(d-3)$-dimensional. The boundaries of these extremal surfaces are $(d-2)$-dimensional so will also sit somewhere on the branes. Their location on the branes also needs to be an extremum. The entanglement entropy is computed by the area of the minimal such surface. 

If we think only of the duality between the CFT$_{d-1}$ and the two $d$-dimensional bulks, this extremization can be thought of as a ``two-sided" extremization which occurs on both sides of the CFT$_{d-1}$. We no longer have a $(d+1)$-dimensional bulk and a minimal surface there, but we can understand what was being captured by that surface. Heuristically, its boundary on the gravitating $d$-branes captures the $d$-dimensional gravitational entropy, while its bulk piece captures the matter entropy of the theory on the $d$-branes (recall again that this division is ambiguous and only the combination is expected to be finite in the ultraviolet theory). Thus, we can dispense with this surface and simply extremize the generalized entropy on the $d$-branes.

\begin{figure}
  \centering
  \includegraphics[scale = .2]{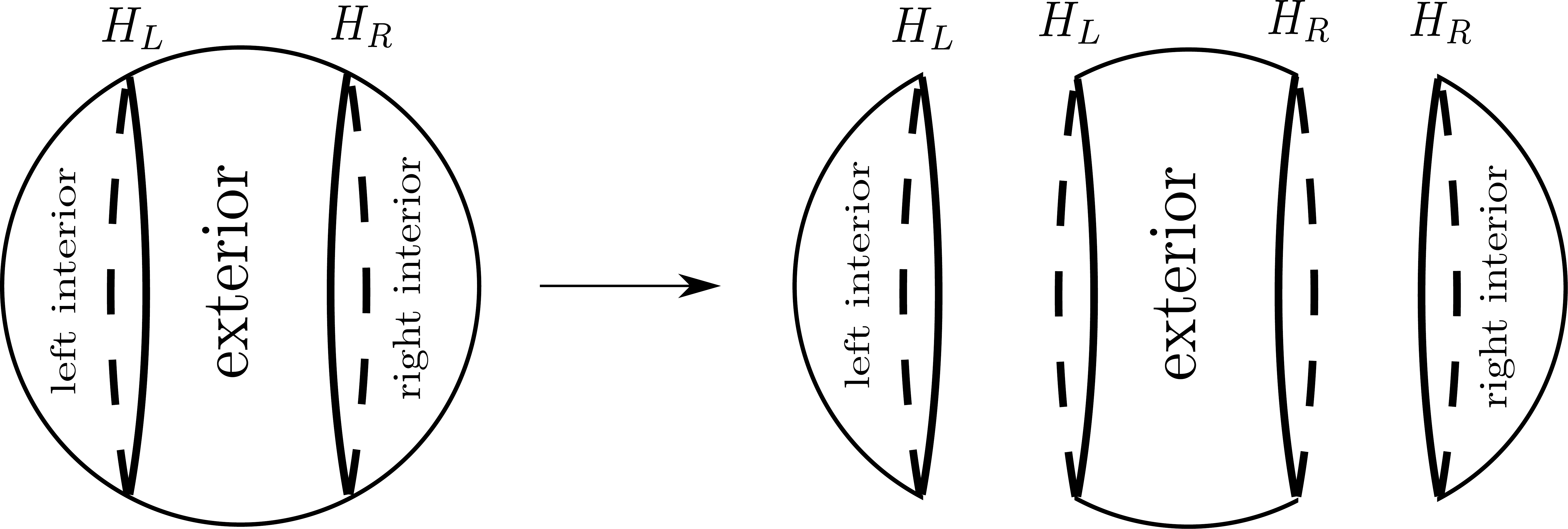}
\caption{A global time slice of de Sitter, with the horizons $H_L$ and $H_R$. The entropy of a subregion $h$ on $H_L \cup H_R$ is computed by an extremal surface which contains $\partial h$ and whose interior includes $h$ and no other piece of $H_L \cup H_R$. This is equivalent to breaking up the sphere into three pieces as in the right image, and doing a single extremization where each surface is anchored to $\partial h$ as in ordinary AdS/CFT. Thinking of the horizons as AdS boundaries makes the homology rule identical to that case. In the classical case with topology fixed as above we can perform three independent extremizations for each region.}\label{dsextreme}
\end{figure}

To compute extremal surfaces in global de Sitter, we therefore have the following minimization problem. We pick a region on the union of the two horizons. We perform a simultaneous ``two-sided" extremization on both horizons with the boundary condition defined by the chosen region. The full entanglement wedge is the region whose boundary is the union of the extremal surfaces found by the procedure above, and which contains the pieces of the horizon whose entropy we are computing. In the classical case with fixed sphere topology, this can be broken up into three independent extremizations: one for each of the two static patch interiors, and one between the two horizons. We will allow our extremal surfaces to have kinks across the horizons. See Figure \ref{dsextreme}. (For a distinct proposal for dealing with the other side of the horizon, see \cite{Murdia:2020iac}.)

\subsection{Physical interpretation}
In wedge holography, the CFT$_{d-1}$ can be thought of as two CFT$_{d-1}$ systems interacting through a shared $d$-dimensional bath region, in the limit that the bath region shrinks to zero size, see Figure \ref{wedge}. We will think of our theory on the de Sitter horizon in a similar way, as two interacting theories (not necessarily CFT$_{d-1}$'s), one nominally describing the interior of the horizon (we will call this the interior theory) and the other nominally describing the exterior (we will call this the exterior theory). The word ``nominally" is used here because determining which theory describes which region of the spacetime is a dynamical question that needs to be answered by computing entanglement wedges, and in any case the two theories here are interacting. The interior and exterior theories are jointly in a mixed state, purified by the partner theories on the other horizon. The exterior theories are connected ``through the bulk," so it is like they are in a deconfined state. The interior theories are not connected through the bulk (for empty de Sitter), so it is like they are in a confined state. The spatial sphere caps off in the interior and does not cap off in the exterior. In the AdS/CFT analogy we can modify the bottom row of Figure \ref{wedge} so that one of the two interacting thermofield double pairs is below its Hawking-Page transition and therefore dual to two independent thermal AdS geometries. This way the geometry caps off at either end, like in the case of de Sitter. Alternatively, the interior theories in the case of de Sitter can (temporarily) deconfine, as represented by the Schwarzschild-de Sitter solution, which has a wormhole going from one interior to the other one. This is more like the bottom row of Figure \ref{wedge}. These horizon theories need not have any spatial locality, so the cuts of the horizon on which we will be anchoring our extremal surfaces need not correspond to spatial divisions of the system. In fact, in the next subsection we will see a ``volume law" for the growth of entropy of a cut on the horizon, indicative of nonlocal behavior. 

The exterior cuts across many horizons, while the interior contains just one horizon -- in the AdS/CFT analogy this would mean the exterior should be described by field theory degrees of freedom while the interior should be described by the matrix degrees of freedom. Studies of the entanglement properties of such matrix degrees of freedom can be found in \cite{Graham:2014iya, Karch:2014pma, Mollabashi:2014qfa, Anous:2019rqb}; our interior extremization is akin to the prescription in these papers.

\subsection{Examples}
Let us pick a few cases to apply our proposal. We will begin by only computing classical extremal surfaces in pure de Sitter spacetime, without any corrections from matter. In this case the extremization problem discussed above can be broken up into three pieces: an extremization that occurs in the interior of each static patch, and an extremization that occurs between the two horizons. 

\subsubsection*{Both horizons: $S(H_L \cup H_R) = 0$}

If we pick the entire boundary region, i.e. both horizons, then the extremization gives the trivial surface. To see it break up into three pieces we have the following: if we extremize a surface homologous to either horizon and contained in its interior, we see that it can shrink and slip off, giving the trivial surface. Similarly, between the two horizons, the two surfaces can coalesce and again give the trivial surface. We get $S = 0$ and the entanglement wedge is the entire spacetime. 

\subsubsection*{One horizon: $S(H_L) = S(H_R) = A/4G$}

If we pick one of the two horizons, then the extremization gives the horizon we chose, with the entanglement wedge being its interior. Let's see this from the three independent extremizations. Extremizing to the chosen horizon's interior gives the trivial surface with the entanglement wedge being the entire interior, while the extremization between horizons gives the degenerate surface which sits exactly on the chosen horizon, thus encoding none of the exterior. (There is a degenerate surface which corresponds to the other horizon, but in such cases we are instructed to pick the surface with the smaller entanglement wedge.) Since no region was chosen on the other horizon no extremization is required to its interior; the entanglement wedge there is trivial. The fact that altogether we encode only one static patch is sensible since -- like with black holes -- having access to only one side restricts the reconstructible region to lay within the corresponding horizon. (We will assume in this section that the horizons are of equal size; see section \ref{exotic} for a discussion of the situation where they are unequal.) An allowed possibility which does not occur is that the chosen horizon's interior extremization leads to an extremal surface in the interior of the \emph{partner} horizon, with the entanglement wedge being everything to the interior of this extremal surface and everything within the original horizon. 

\subsubsection*{Subregion on one horizon: $S(h\subset H_L)$}

Subregions on the horizon are a little more subtle. Say we choose a subregion on just one of the horizons. There are two extremal surfaces, one with area equal to the area of the surface itself and one with area equal to the area of the surface's complement (restricted to the horizon).\footnote{These extremal surfaces do not live at fixed global time even though they will be drawn as if they do for simplicity, see Appendix \ref{horgeo} for a discussion.} The smaller of the two is clearly the minimal one for the interior extremization. %(In the case of exactly half the horizon there is an infinite number of degenerate extremal surfaces, see Appendix \ref{app}.)

\begin{figure}
  \centering
  \includegraphics[scale = .6]{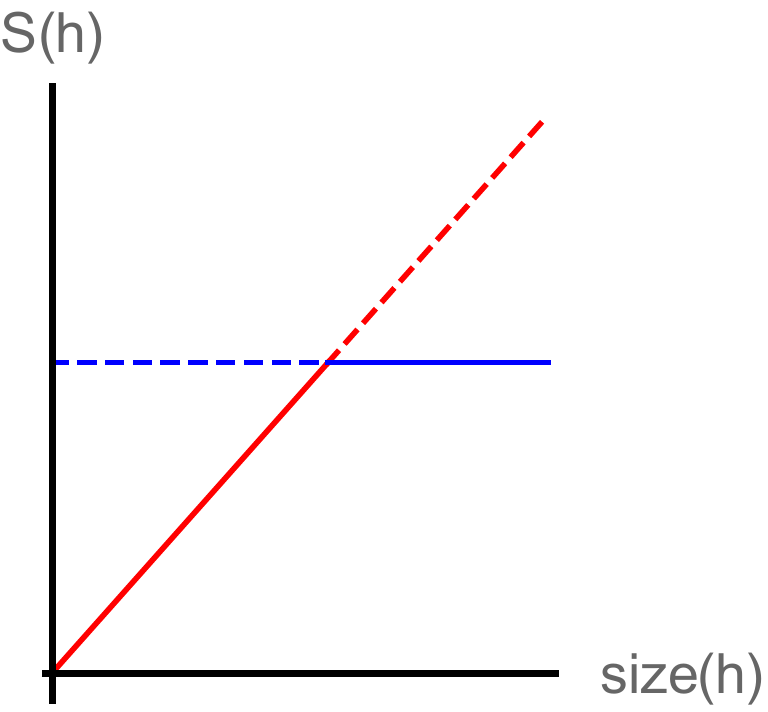} \qquad\quad  \includegraphics[scale = .25]{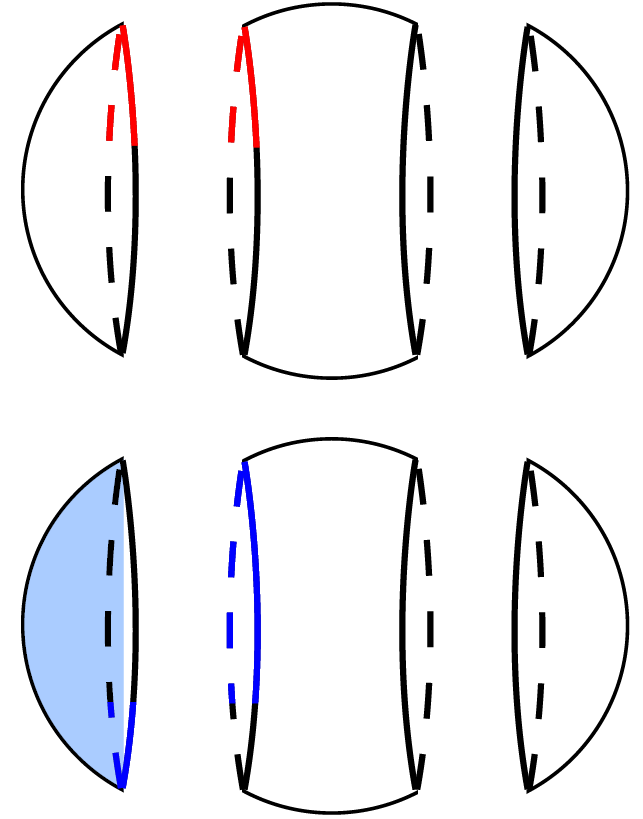}
\caption{Left: There are two extremal surfaces for the entropy of a subregion $h$ of one horizon, marked as red and blue above. The minimal one exhibits a transition: the entropy of a subregion grows as twice its size (solid red curve) until the halfway point, where it saturates at the horizon area (solid blue curve) and the entanglement wedge becomes the interior of the horizon. This transition will be smoothened out by anchoring at the stretched horizon. Right: the red saddle and blue saddle drawn for configurations where they are minimal, with the entanglement wedge shaded. (These extremal surfaces do \emph{not} live at fixed global time -- they are drawn that way above for simplicity, see Appendix \ref{horgeo} for a discussion.) This is reminiscent of the island transition in an eternal black hole.}\label{hawkingpage}
\end{figure}

This is not quite true for the exterior extremization due to the homology constraint; the appropriate extremal surface will always live on the horizon and have area equal to the area of the region we are considering. The surface with area equal to the area of the complement of our region on the horizon violates the homology constraint since the entanglement wedge will have the partner horizon as a boundary. This leads to the following intriguing situation. For a subset of the horizon $h$ which is smaller than half, none of the bulk interior or exterior is encoded, and $S = 2\, \text{Area}(h)$. This is a ``volume law" growth of the entropy, indicative of nonlocality of the dual theory. This threatens to violate unitarity when $h$ is half the system size and $S = A/4G$ for $A$ the total horizon area, but at this point there is a transition. The exterior extremization continues to give Area$(h)$ while the interior extremization flips to Area$(\bar{h})$. The entanglement wedge is now the entire interior (to see this we can regulate the horizon into the stretched horizon and anchor our surfaces there), and the entropy stays pegged at $S = A/4G$. This transition at half the system size and the subsequent encoding of the interior seems tantalizingly like the island transition, see Figure \ref{hawkingpage}.

We can smoothen this transition by putting our anchoring surface at the stretched horizon. This breaks the degeneracy and for an interior extremization of a subregion on the horizon produces entanglement wedges of increasingly bigger size as the subregion size is increased.

\subsubsection*{Subregions on both horizons and a Hartman-Maldacena transition: $S(h_L \cup h_R)$}
\begin{figure}
  \centering
  \includegraphics[scale = .2]{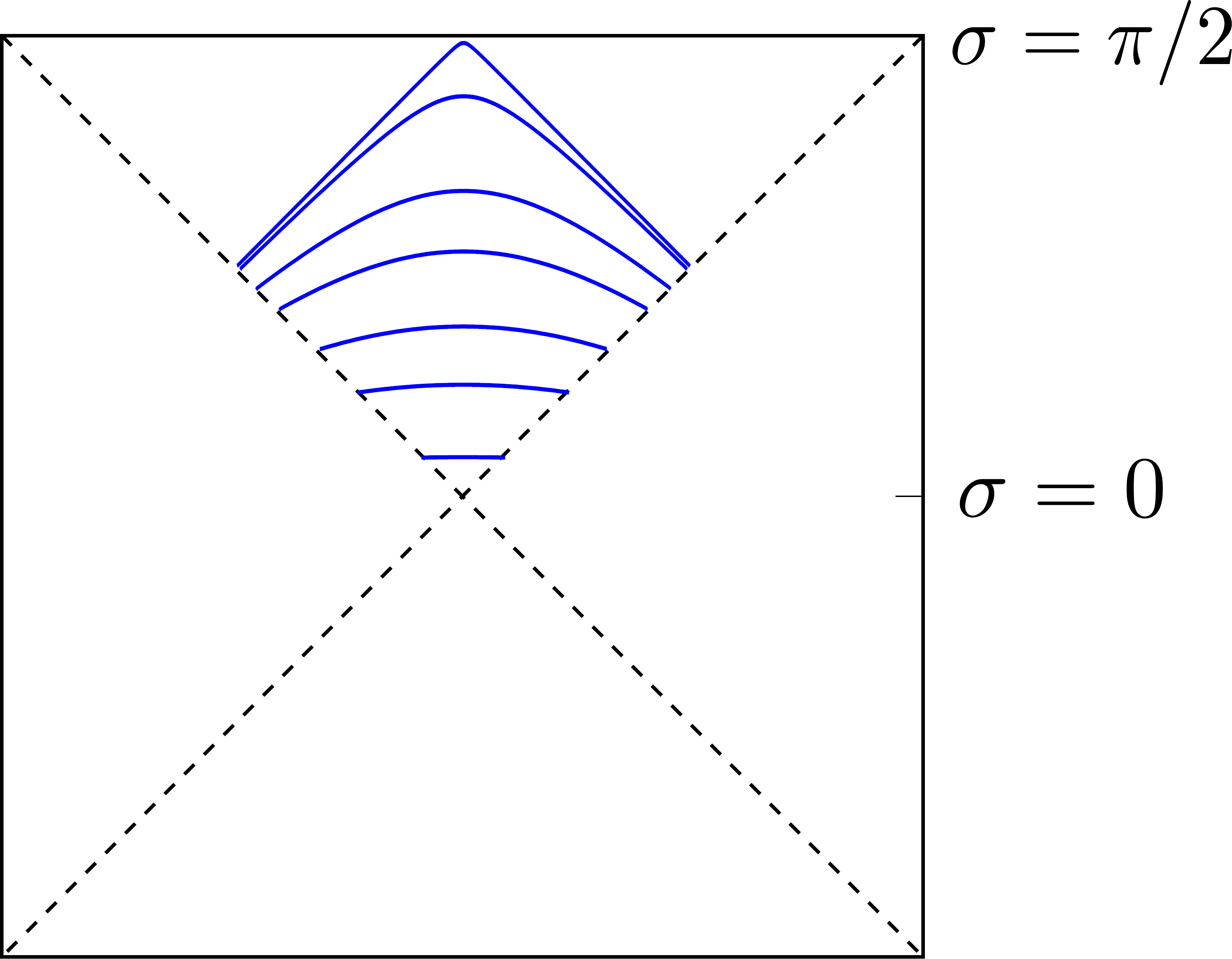}
\caption{The extremal surfaces which connect a piece of one horizon to the same piece of the other horizon. They cease to exist past $\s = \pi/4$, before which a transition to the disconnected surface which sits on the two horizons occurs; this is like the Hartman-Maldacena transition in AdS/CFT. The area of this saddle diverges as $\s \rightarrow \pi/4$ for all dS$_{d>3}$. For dS$_3$ if we pick the same interval on each horizon the length of this saddle approaches a single horizon length $2\pi$.}\label{hmtransition}
\end{figure}

We can also consider similar subregions on each horizon.\footnote{This calculation was suggested by Leonard Susskind.} The interior extremizations remain the same as we discussed above, although now we have a competition between three saddles for the exterior extremization. The first two saddles are disconnected and live on the horizon, with area equal to the area of the region we are considering or its complement (restricted to the horizon). Any time the regions on the two horizons have a summed area $A>A_{dS}$, then the outer extremization picks the complements on the horizon, with area $2A_{dS}- A< A_{dS}$. In this case the entanglement wedge is the entire inflating region, otherwise it is none of the inflating region. The third saddle is connected across the inflating region, represented in Figure \ref{hmtransition}. At early times the connected saddle will dominate, due to the short distance between the two horizons. The entropy implied by this saddle grows without bound for dS$_{d>3}$, threatening the bound set by finite quantum systems with exp$(A/4G)$ states. A transition to the disconnected saddle occurs before any violation, and the entropy becomes time-independent. This is a de Sitter version of the Hartman-Maldacena transition \cite{Hartman:2013qma}. Computational details can be found in Appendix \ref{app}. 

Notice that the exterior extremization is necessary not only to answer the question of how the inflating region is encoded, but also to get $S = A/4G$ in the limit where we take one entire horizon as in the above example.

\subsubsection*{Necessity of joint extremization}
The example of de Sitter-Schwarzschild simply illustrates why our proposed extremization does not generically break up into separate extremizations. In particular, the two static patch interiors are now connected by a wormhole. As usual, if we compute the full entropy, i.e. that of both cosmological horizons, we expect to get $S = 0$. If we performed the interior extremizations independently then each extremization would land on the black hole horizon and give $S = A_{BH}/4$ for $A_{BH}$ the area of the black hole. Performing a joint extremization shows that the extremal surfaces can go through the black hole wormhole and coalesce, giving $S = 0$ and an entanglement wedge which covers the entire spacetime. So we see that connections between regions, either through classical bridges or quantum entanglement (this will be introduced in Section \ref{qesrule}), necessitates a joint extremization.

 If we compute the entropy of one of the two cosmological horizons, we find $S = A_{BH}/4 + A_{CH}/4$, where $A_{CH}$ is the area of the cosmological horizon. 

As in the pure de Sitter case, the exterior extremizations for symmetric regions $h_L \cup h_R$ in this case have connected surfaces which cut across the inflating region, as long as we are not at late times. The interior extremization for Schwarzschild-de Sitter can also have connected surfaces which cut across the black hole interior, analogous to the Hartman-Maldacena surface for eternal black holes in AdS/CFT. In certain regimes -- for example big black holes close to the Nariai bound at early times -- the dominant surface will cut across the inflating de Sitter region \emph{and} the crunching black hole region. 

The quantum-mechanical problem will have to deal with the instability of de Sitter-Schwarzschild black holes, which will greatly change the nature of the problem. 

\subsection{Encoding the exterior with one side}\label{exotic}
Most of the solutions we consider in dS, if starting from one static patch and analytically extended, lead to equal-size horizons. For example if you put a particle in one static patch to make the horizon smaller, the extended Schwarzschild solution implies a particle in the other static patch which makes its horizon smaller as well. One way to think about this is that one needs to recover vanishing energy on compact slices, and the time translation Killing vector runs in opposite directions in the two patches. If we do not insist on this symmetry, however, we can have solutions with horizons of different sizes. In particular we can imagine sending some matter into the past lightcone of one observer but not her antipodal partner's past lightcone.

If such solutions are allowed and consistent in the full theory, then our prescription above implies that the theory living on the bigger horizon, say $H_L$, can encode the entire exterior by itself. This is simply because if we do the exterior extremization for $H_L$, the extremal surface will want to sit on the smaller horizon $H_R$ and the entire exterior will be encoded. For smaller regions $h \subset H_L$, the exterior extremization can give the ``complement" surface given by $H_R$ and a region on the left horizon with area equal to the area of $\bar{h}$.

Notice that in the thermofield double black hole analogy this type of encoding is impossible: none of the interior of the black hole is encoded if we restrict to just one of the two CFT's. The interior is instead encoded in the entanglement between the two. This means that the horizon theories in the case of de Sitter spacetime must be interacting. Notice that even if the bigger horizon encodes the entire exterior, its reconstruction is very likely complex. For $H_L$ the true minimal surface for the extremization of $H_R$, we have $H_R$ as a subleading minimal surface and the waist of the sphere inbetween as a maximal surface. This leads to a Python's lunch situation, where \emph{everything} in the exterior is exponentially complex to encode with only the data of the bigger horizon theory. This is like the island that appears inside an evaporating black hole. Interactions between the quantum system describing the black hole and a bath where the Hawking radiation escapes allow the entanglement wedge of the bath to include the interior of the black hole, although the reconstruction of the interior is exponentially more complex with only access to the Hawking radiation. Here interactions between the two horizon theories can allow one of them to encode their shared exterior, although the reconstruction of the exterior is exponentially more complex with only access to one of the two horizon theories.

\subsection{Bulk reconstruction}
For entanglement wedges to correspond to regions that are reconstructible from the dual theory's degrees of freedom, there are two things we will want to be true: (a) entanglement wedges of smaller Hilbert spaces in the boundary theory are contained within the entanglement wedges of larger Hilbert spaces, (b) causal wedges, defined as the intersection of the past and future of the domain of dependence of the horizon region $A$, are always contained within entanglement wedges. 

Although it is not clear that (b) need be true in our situation, it is nevertheless trivially true. The causal wedge of any piece of the horizon reduces to the piece itself, so it has no extent into the interior or exterior. So causal wedges will at least be weakly contained within entanglement wedges, and as we saw above generically we have nontrivial entanglement wedges which can probe the interior and exterior. 

Given that the horizons make up suitable holographic screens for this spacetime, entanglement wedge nesting is expected to hold for extremal surfaces anchored to them \cite{Sanches:2016sxy}. In particular our interior extremization obeys entanglement wedge nesting (to make it nontrivial we should regulate to the stretched horizon so that the extremal surfaces do not simply sit on the horizon but instead sweep out an increasingly larger portion of the interior). So let's focus on the more peculiar exterior extremization. In the simple case of $S(h_L)$ the extremal surface and entanglement wedge are just $h_L$, so nesting is trivially satisfied. In the case of $S(h_L \cup h_R)$ at some fixed global time $\s > 0$ and small region size, we begin with the entanglement wedge as $h_L \cup h_R$. As we grow the region size we can transition to the connected surface which cuts through the inflating region. Let's say $h_L$ and $h_R$ are symmetric regions on the horizons with azimuthal angle  $\theta \in [(\pi - \theta_c)/2,(\pi+\theta_c)/2]$. The area of the connected and disconnected surfaces both grow as $\sin^{d-2}\theta_c$, so the connected surface will remain dominant until the area of $\overline{h_L}$ becomes smaller than the area of the connected surface. At this point we transition to the complement region $\overline{h_L} \cup \overline{h_R}$, and the entanglement wedge is the entire exterior. 

\subsection{Quantum extremal surface prescription}\label{qesrule}
The full prescription for computing the entropy  is proposed to be an extremization of the generalized entropy
\be
S_{\text{gen}} = \f{A}{4G} + S_{\text{matter}}.
\ee
This becomes interesting in situations where we have entangled particles, with one in the interior of a static patch and its partner in the exterior. An example where the partner is in the inflating region can be prepared simply by entangling two qubits and throwing one past the horizon. An example where the partner is in the neighboring static patch can be prepared by a path integral which inserts entangled qubits, one in each static patch. (The Schwarzschild-de Sitter black hole is a classical version of this). 

As a simple application, we can consider an observer within one of the static patches creating a large black hole out of pure state matter. We take $\ell_{dS} \gg r_{BH} \gg \ell_{Pl}$, i.e. a semiclassically large black hole which is much smaller than the horizon scale. We wait until a time well after the Page time and collect all the Hawking radiation from the black hole. This Hawking radiation encodes the interior of the black hole, and we can now throw it past the cosmological horizon such that the exterior Hawking quanta which encode the interior of the black hole have fallen past the cosmological horizon. In such a situation, the exterior extremizations are going to be important in recovering the interior of the black hole, since the exterior behaves as a ``bath" for the interior. 

\section{Discussion}
We have investigated whether there is evidence for a cosmological central dogma and highlighted a few puzzles with such a picture, including the one of redundant encoding (where completely distinct regions can encode the same piece of spacetime) and the nature of the minimax bifurcation surface in de Sitter space. The inclusion of this surface leads to violations of entanglement wedge nesting in a conventional extremization problem. We view this as a crucial difference with the case of the eternal black hole, where the bifurcation surface is instead a maximin surface and gives sensible results. In particular, this makes one question the Gibbons-Hawking entropy as being related to a microscopic count of states. An alternative role for the de Sitter bifurcation surface may be as a measure of complexity instead of entropy. In particular, de Sitter space provides the most pristine example of a Python's lunch: every time slice is a sphere, the constrictions are of vanishing size and the lunch is the entire spatial manifold. 

This problem led us to consider anchoring extremal surfaces to the horizon itself, which allowed the horizon to act as a maximin surface since its location was pinned. This prescription reproduced some features expected of a quantum description of de Sitter space, including vanishing total entropy and an entropy of $A/4G$ when restricting to a single static patch. It also exhibited some novel transitions: when considering a subregion on one horizon we had a transition where the entanglement wedge went from covering none of the interior to covering all of the interior as we grew the subregion, and when considering subregions on both horizons we had a de Sitter version of the Hartman-Maldacena transition as we evolved in time. Both these transitions save us from violating the entanglement bound set by assuming the horizon theories are finite quantum systems with exp$(A/4G)$ states. The most exotic feature seemed to be the ability of one of the two horizon theories to encode the entire exterior, which must mean that the two horizon theories are interacting. This seems natural from the perspective that perturbations make the de Sitter Penrose diagram grow taller \cite{Gao:2001ut, Leblond:2002ns}. Notice that the extremal surface moving from the larger horizon to the smaller horizon also keeps us from violating the entanglement bound just discussed.

The nature of each horizon theory is unclear, although we tried to draw an analogy with wedge holography in AdS/CFT. The description of the exterior is likely to follow more conventional ideas from AdS/CFT since it probes super-horizon physics, whereas for the interior the matrix degrees of freedom will have to play a crucial role since it must probe sub-horizon physics, something that is not well understood even in AdS/CFT. 

Altogether, this is suggestive of a trivial global Hilbert space (unpinned extremal surfaces always want to vanish if gravity is fluctuating everywhere) and a rich local picture once we try to split the space into pieces, which in our construction was done by pinning extremal surfaces to the horizons. 

If an observer can decode what is beyond their horizon, then distinct observers who do not interact can independently decode beyond their horizons, which overlap with one another. This is akin to entanglement wedges of disjoint regions overlapping in AdS/CFT, which would lead to various paradoxes. As formulated, the prescription in this paper evades this conceptual puzzle since it has complementary recovery built in. However, the dynamical nature of the horizon, its dependence on a choice of observer (see Appendix \ref{observdep} for some discussion on the observer-dependence of horizons in the black hole and cosmological cases), and the lack of a gravitational decoupling limit remain as important challenges for developing this picture. 

%An important conceptual puzzle is the observer-dependence of this picture. Fixing to a pair of static patches is like fixing to a pair of observers, but this choice was arbitrary. In particular, there are an infinity of observers in de Sitter space with different cosmological event horizons, contrasting with the case of a black hole. (See Appendix \ref{observdep} for some discussion on this.) Beyond working out the nature of the horizon theories, it is also important to understand how they relate to horizon theories of other pairs of horizons that may be chosen. This is related to the issue of redundant encoding. 

\bigskip\bigskip\bigskip

\noindent \textbf{Acknowledgments} I would like to thank Leonard Susskind for useful conversations and sharing a draft of \cite{Susskind:2021dfc} on August 30. I would like to thank Raghu Mahajan for collaboration on related subjects and many useful discussions/comments on a draft. I would also like to thank Dionysios Anninos, Tom Hartman, and Adam Levine for useful conversations. This work was supported by the Simons Foundation It from Qubit collaboration (385592) and the DOE QuantISED grant DESC0020360.

\appendix
\section{Extremal surfaces}\label{app}
\begin{figure}
  \centering
  \includegraphics[scale = .15]{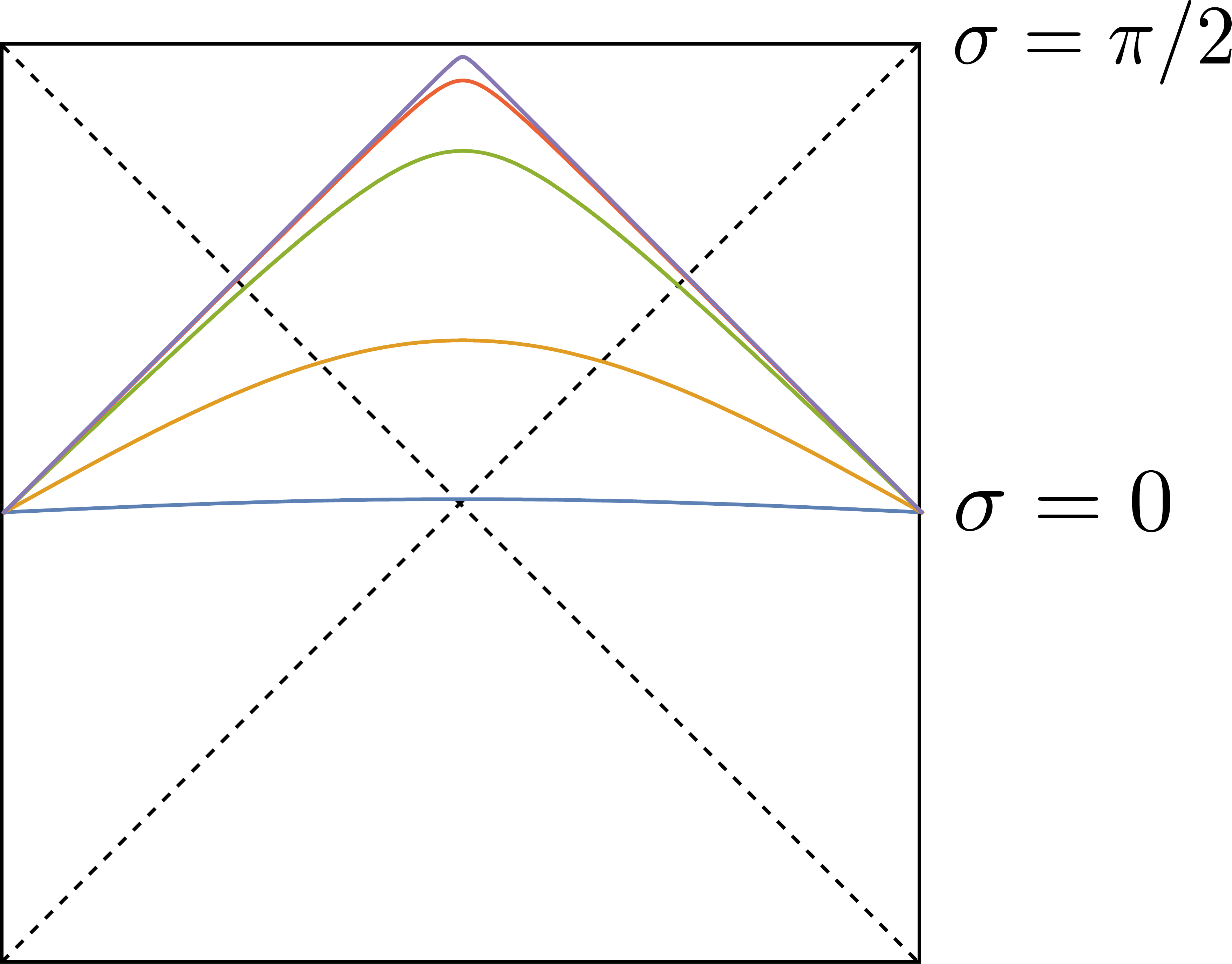}
\caption{The geodesics in dS$_3$ which go from one horizon to the partner horizon. They are shown extended into the interiors of the static patches. There are no real spacelike geodesics connecting horizons beyond $\s = \pi/4$, where they approach a null geodesic. The full geodesics from the static patch origin at $\s = 0$ to its antipodal point have equal length and are part of the infinite family of degenerate geodesics connecting antipodal points on de Sitter space.}\label{geodesicsDS}
\end{figure}

Here we compute the extremal surfaces relevant for the exterior extremization in pure de Sitter space, i.e. the extremal surfaces which connect one horizon to the partner horizon. We will begin with the case of pure dS$_3$:
\be
ds^2 = -(1-r^2)dt^2 + \f{dr^2}{1-r^2} + r^2 d\phi^2
\ee
We will be using this metric for $r>1$, where $r$ is the timelike coordinate and $t$ is the spacelike one. We pick symmetric regions on the two horizons, $\phi \in [\phi_1, \phi_2]$. The surface we are looking for will connect $\phi_1$ on one horizon to $\phi_1$ on the other horizon (without any winding around the $\phi$ circle), and similarly for $\phi_2$. This means we can set $\dot{\phi} = 0$, since otherwise the final $\phi$ value will be different than the initial value. The value of $\phi$ is immaterial, and we parameterize our surface as $r(t)$ to write the length functional:
\be
\text{Length} = \int dt L[r,\dot{r}] = \int dt \sqrt{-(1-r^2) + \f{\dot{r}^2}{1-r^2}}
\ee
This immediately gives the conserved quantity $\f{\partial L}{\partial \dot{r}} \dot{r}-L = E$, which we can use to solve for $\dot{r}$ and write
\be
\text{Length} = \int \f{dr}{\dot{r}}L[r]\,.
\ee
By the symmetry of our problem the turning point of the geodesic will be where $\dot{r} = 0$, which is at $r = \sqrt{1+E^2}$. Integrating from $r=1$ to this point gives us half of one geodesic, and the total length of the two geodesics is therefore four times this result:
\be
\text{Length} = 4 \tan^{-1} E\,.
\ee
The length goes from $0$ at $E=0$ to a maximum of $2\pi$ as $E \rightarrow \infty$. Notice that if we integrated the length all the way down to $r = 0$, we would get $2\pi$, independent of $E$. This shows that these geodesics are simply (pieces of) the infinite family of degenerate geodesics which connect the static patch origin $r=0$ at $t=0$ to its antipodal point, analogous to the infinite family of degenerate geodesics connecting antipodal points on the sphere.\footnote{Recall the following construction for obtaining the geodesics of dS$_3$. Embed the dS$_3$ hyperboloid into $\mathbb{R}^{(3,1)}$ with the waist surrounding the origin, and for any two points on dS$_3$ obtain the geodesic by the intersection of dS$_3$ with a plane passing through the two points and the origin of $\mathbb{R}^{(3,1)}$. Antipodal points at the waist have an infinite family of such planes since the three points in question lie on a line.} In the inflating region these geodesics become more and more null as they intersect the horizon further in the future, approaching a null geodesic which bounces of $\mathcal{I}^+$ before returning to the antipodal point. In global coordinates, 
\be
ds^2 = \f{-d\s^2 + d\theta^2 + \sin^2 \theta d\phi^2}{\cos^2\s}\,,\qquad \s \in [-\pi/2,\pi/2]\,,
\ee
the geodesic approaches being exactly null when intersecting the horizon at $\s = \pi/4$, $\theta = \pi/4$. For $\s>\pi/4$ there are no real spacelike geodesics connecting equal-$\phi$ points on the horizon. This is shown in figure \ref{geodesicsDS}. The lack of extremal surfaces past $\s = \pi/4$ persists in higher dimensions. Similar behavior in two dimensions has implications for complexity, as recently explored in \cite{Chapman:2021eyy}. 

Altogether we have a competition between the geodesics which sit on the horizon and the ones which cut across the inflating region. At $\s = 0$ we begin with the ones which cut across the inflating region since they have zero length. However, their length increases in time, instigating a transition to the geodesics which sit on the horizon and have length min$\{2\Delta \phi, 2(2\pi - \Delta\phi)\}$ for $\Delta \phi = \phi_2 - \phi_1$. The minimization simply picks either the region or its complement (restricted to the horizon), whichever is smaller. The case $\Delta \phi = \pi$ is somewhat degenerate, since at time $\s = \pi/4$ all three surfaces have equal length. 

\subsection*{Higher dimensions}
We can look for codimension-two extremal surfaces connecting the horizons in higher dimensions. We will also slightly generalize the problem and consider the de Sitter-Schwarzschild solution:
\be\label{connected}
ds^2 = -\left(1-r^2 - \f{\m}{r^{d-2}}\right)+\f{dr^2}{1-r^2 - \f{\m}{r^{d-2}}} +r^2d\Omega_{d-1}^2\,.
\ee
This solution is quantum-mechanically unstable: outside of the Nariai limit the black hole and the cosmological horizons are at different temperatures and therefore out of thermal equilibrium. We will only focus on the classical problem and ignore Hawking radiation. 

In dS$_3$ there are no black hole solutions, so the family above corresponds instead to conical deficits (for $0<\m<1$) sourced by a positive mass particle at the origin of the static patch and its antipodal partner. This case is trivially related to the one with $\m=0$, and the length is found to be $4 \tan^{-1}\f{E}{\sqrt{1-\m}}$; by rescaling the integration constant $E$ this is the same as before. This is because the geodesics, like the geometry, are obtained by a quotient of the $\m = 0$ spacetime where we take $\phi \sim \phi + 2\pi \sqrt{1-\m}$. Thus its only effect is going to be on global properties. Indeed, since the horizon is now smaller, the disconnected geodesic has a maximum value of $2\pi \a < 2\pi$. The connected geodesics are unaffected and still have maximum length $2\pi$ since they are $\phi$-independent. In this sense the transition from disconnected to connected surface happens more quickly as $\m$ is increased. Notice that a conical surplus, which corresponds to a negative mass particle, would lead to the uncomfortable situation where the connected surface can cease to exist before the transition to the disconnected surface happens. 

For $d>2$ we have
\be
\text{Area} = 2\text{ Area}(S^{d-2}) \sin^{d-2}\theta_1 \int dt\, r^{d-2} \sqrt{-\left(1-r^2-\f{\m}{r^{d-2}}\right)+\f{\dot{r}^2}{1-r^2-\f{\m}{r^{d-2}}}}
\ee
for $\theta_1$ the angle of the bounding $S^{d-3}$ of the region on the horizon $S^{d-2}$ whose entropy we are trying to compute. The factor of two comes from there being two bounding $S^{d-3}$ spheres, which we have oriented to be equal-sized. The structure of the extremal surfaces is similar to the geodesics in dS$_3$, except in this case as the surface approaches $\mathcal{I}^+$ and becomes null its area diverges. Heuristically, this is because there is a transverse sphere coming along for the ride, but its ride now includes regions arbitrarily close to $\mathcal{I}^+$, which are infinitely big. Past this extreme null surface we have no extremal surface connecting the patches on the horizons. The black hole geometry for $d>2$ is not a quotient, but  still acts to pull in the horizon. Thus the disconnected surface can be favored by regions that are sufficiently small or sufficiently big. 

\section{Observer dependence}\label{observdep}
\begin{figure}
  \centering
 \hspace{-5mm} \includegraphics[scale = 2]{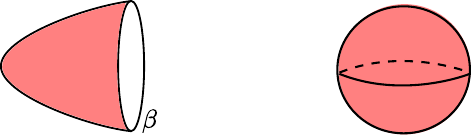}
\caption{Left: the thermal circle $\b$ caps off smoothly in the Euclidean black hole spacetime, at a special point dictated by the time of the asymptotic exterior observer. Right: in Euclidean de Sitter every point is equivalent, although we can write it as an $S^1 \times S^{d-2}$ fibered over an interval in an infinite number of ways, with the $S^1$ degenerating to zero size at any point on the sphere that we wish. This freedom corresponds to the choice of observer.}\label{BHcigar}
\end{figure}

A natural complaint against anchoring extremal surfaces on cosmological horizons is the observer-dependence of the cosmological horizon. One way to distinguish this from the black hole case is to think of the Euclidean picture. The Euclidean black hole can be represented as a cigar, for example in flat spacetime we have
\be
ds^2 = \left(1-\f{\m}{r^{d-3}}\right)d\t^2 + \f{dr^2}{1-\f{\m}{r^{d-3}}} + r^2 d\Omega_{d-2}^2\,,
\ee
 which can be represented as in Figure \ref{BHcigar}. The horizon is a special point, the tip of the cigar. Euclidean de Sitter space is a sphere, for which all points are on equal footing.\footnote{This was emphasized to me by Dionysios Anninos.} Thus, picking a pair of horizons should be thought of as some sort of gauge-fixing: we pick an observer in a patch and write the metric according to her time $t$:
\be
ds^2 = -(1-r^2)dt^2 + \f{dr^2}{1-r^2} + r^2 d\Omega_{d-2}^2\,.
\ee
When we continue this metric to Euclidean time $t \rightarrow i\tau$, we see that we need to identify $\t \sim \t + 2\pi$ to maintain a smooth metric at $r=1$. We then have the $S^d$ metric written as $S^1 \times S^{d-2}$ fibered over an interval. The thermal circle degenerates at $r=1$ (while the sphere degenerates at $r=0$). This now mimics the black hole case, and in fact the two are more similar than it seems. There is of course an observer-dependence to the black hole horizon as well; somebody who jumps in has a different horizon. So we are implicitly fixing to the exterior observer, usually one at asymptotic infinity, and using her time to write the metric of the black hole. But observers exterior to the black hole horizon having the same horizon is the same as observers interior to a cosmic horizon having the same horizon. The difference of course is that there are an infinity of observers in de Sitter with distinct horizons, and that -- if we can trust the classical picture which persists to $\mathcal{I}^+$ -- there is no way to evade being within a horizon. 

If quantum mechanics on the scale of global de Sitter makes sense, it is reasonable that it has a one-dimensional Hilbert space. One way to think about this is so that there is no factorization problem \cite{Coleman:1988cy, Giddings:1988cx, Marolf:2020xie, McNamara:2020uza, Anous:2020lka}. In that case we are really speaking about the baby universe Hilbert space, although it is plausible that it being one-dimensional implies the same about a single de Sitter universe. 

It is important for the consistency of this picture to recall that the cosmological horizon entropy is invisible in global coordinates:
\be
ds^2 = -dt^2 + \cosh^2 t d\Omega_{d-3}^2 \rightarrow d\t^2 + \cos^2 \t\, d\Omega_{d-3}^2\,.
\ee
In this case there is no periodicity to the time coordinate $\t$. If the Hilbert space is one-dimensional, then the entropies and Hilbert spaces we are discussing come from splitting up the sphere into pieces, as we saw above when we fixed to an observer. An example of this is a topological theory on a spatial sphere, e.g. Chern-Simons theory on $S^2$. Such a theory has no local operators, therefore no states on the sphere. But if we split the sphere into two pieces, the theory comes to life: edge modes imply a nontrivial entropy between subregions.\footnote{Gravity is even more topological than topological quantum field theories. For example, Chern-Simons theory on the torus $\mathbb{T}^2$ has a rich Hilbert space, since the states on this manifold map to line operators. We can change the topology of de Sitter space by considering e.g. the Schwarzschild-de Sitter solution, but we still expect it to have a trivial Hilbert space.} See \cite{Anninos:2020hfj} for a recent extensive analysis of this picture. 
%Global de Sitter space is a cold, dead place. It is only when we split the Hilbert space between observers and antipodal observers that it comes to life. 

\section{Extremal surfaces on the horizon}\label{horgeo}
In this section we discuss the extremal surfaces which sit on the horizon. The subtlety we would like to explain is that they do not sit at a constant time along the horizon, i.e. they are not at a constant global time. 

\begin{figure}
  \centering
 \hspace{-5mm} \includegraphics[scale = .18]{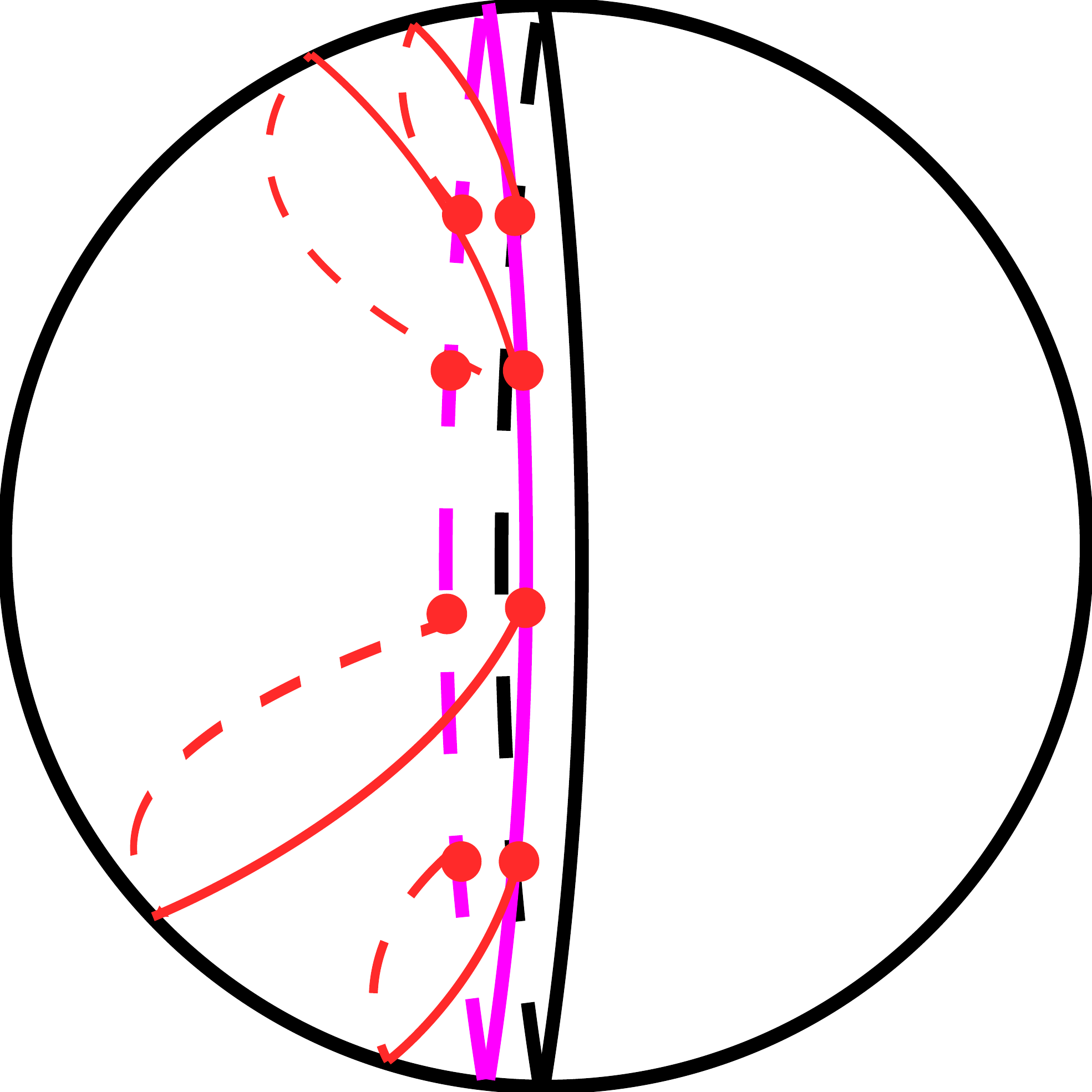}
\caption{dS$_3$ at time $t=0$ is given by an $S^2$ with the bifurcate horizon as the great circle, represented in black. We regulate it into a stretched horizon, drawn in purple, on which we anchor extremal surfaces. These surfaces reach into the bulk, although as the stretched horizon approaches the bifurcate horizon the surfaces remain exactly on the horizon. The extremal surfaces anchored to the stretched horizon can be translated to finite time $t>0$ using the static patch time-translation Killing vector, which puts them very close to the horizon at later times.}\label{stretched}
\end{figure}

One way to think about these surfaces is as a limit of the extremal surfaces at $ t= 0$ when we slightly regulate the horizon. In this case by time-reflection symmetry the extremal surface will live on the $t=0$ slice, which is a sphere. The extremal surfaces here are just the ``great spheres," for example for dS$_3$ we simply have great circles which sweep out bigger and bigger regions of the interior and connect intervals on the regulated horizon, see Figure \ref{stretched} (if we didn't regulate and anchored on the horizon at $t=0$, then the great circles would live exactly on the horizon). We can use the time-translation Killing vector of the static patch to move these extremal surfaces to finite time. The closer we pick our regulated horizon to the true horizon, the closer these geodesics will get to approximating geodesics when we sit exactly on the horizon at a finite global time $\s>0$. But notice that the time-translation symmetry says that the extremal surface will live at a fixed static patch time $t$, which becomes increasingly null the further in time we go. So the surface will have to move in global time $\s$. In particular, in the limit of anchoring to the actual horizon at finite $\s>0$ the extremal surface should move up and down the event horizon. We will now show this for the case of dS$_3$.

The geodesics of de Sitter space can be obtained simply in the embedding formalism. We write a unit-radius dS$_d$ as the hyperboloid
\be
-X_0^2 + X_i^2 = 1\,,\qquad i = 1, \dots, d
\ee
embedded in $\mathbb{R}^{(1,d)}$ with metric $ds^2 = -dX_0^2 + dX_i^2$. The geodesics are obtained by intersecting 2-planes which pass through the origin of $\mathbb{R}^{(1,d)}$ and the two points on dS$_d$ which we want to connect. Restricting now to dS$_3$, we pick some Minkowski time $X_0 = T$, which has horizons at $X_1 = \pm T$.\footnote{Consider global coordinates $X_0 = \tan \s$, $X_1 = \f{\sin \theta}{\cos \s}$, $X_i = \f{\cos \theta \omega_i}{\cos \s}$ for $\sum_i \omega_i^2 = 1$, giving $ds^2 = (-d\s^2 + \cos^2 \theta \,d\Omega^2)/\cos^2 \s$. In this metric the horizons are at $\s = \pm \theta$, i.e. $X_0 = \pm X_1$.} Let's focus on the horizon at $X_1 = T$, which by the embedding equation above corresponds to $X_2^2 + X_3^2 =1$. We use the translational symmetry along the horizon to fix one point to $X_2 = 1, X_3 = 0$ and parameterize the other point as $X_2 = \cos \phi , X_3  \sin \phi$. So we want the plane which connects the three points
\be
(0,0,0,0)\,,\qquad (T, T, \ell, 0), \qquad (T,T,  \cos \phi, \sin \phi). 
\ee
This is parameterized by 
\be\label{horslice}
t_1 (T, T, 1, 0) + t_2 (T,T, \cos \phi, \sin \phi)\,.
\ee
The origin is given by $(t_1, t_2) = (0,0)$ while the points on the horizon are given by $(1,0)$ and $(0,1)$. The rest of the geodesic is given by intersecting this plane with the hyperboloid, which gives the constraint
\be\label{horcons}
(t_1+t_2\cos \phi)^2 +t_2^2 \sin^2 \phi =1 \implies t_1^2 + 2 t_1 t_2 \cos \phi + t_2^2 = 1\,. %t_1=- \alpha t_2 + \sqrt{1+t_2^2 (\a^2-1)}\,. 
\ee
There is a continuous path satisfying this constraint equation which goes from $(t_1, t_2) = (1,0)$ to $(0,1)$, which shows that there is a geodesic on the dS$_3$ hyperboloid connecting the two points. Its length equals the length of the region of the horizon chosen, since the null extent does not contribute. Notice that \eqref{horslice} already tells us that the rest of the geodesic has to be on the horizon, since all points satisfy $X_0 = X_1$, while \eqref{horcons} tells us that the only points on the geodesic which are at time $X_0 = T$ are the endpoints; the rest of the geodesic moves up in time $X_0$ (and in space $X_1$) since $t_1^2 + 2  t_1 t_2 \cos\phi+ t_2^2 = 1 \implies t_1 + t_2 > 1$ unless one of $t_1$, $t_2$ equals zero. This is consistent with the argument at the beginning of this section which shifted the $t=0$ extremal surfaces using the static patch time translation generator.

\footnotesize
\bibliographystyle{ourbst}
\bibliography{dScentraldogmaDraftV3}

\end{document}